\documentclass[a4paper,fleqn,usenatbib]{mnras}

\usepackage[T1]{fontenc}
\usepackage{ae,aecompl}

\usepackage{graphicx}	
\usepackage{amsmath}	
\usepackage{subfig}
\usepackage{enumitem}
\usepackage{color}
\usepackage[dvipsnames]{xcolor}

\usepackage{newtxtext,newtxmath}
\usepackage{balance}

\newcommand{\hone}{\textsc{Hi}}
\usepackage{hyperref}

\usepackage{xcolor}

\title[High resolution full-sky spectral index simulation]{Simulating a full-sky high resolution Galactic synchrotron spectral index map using neural networks}

\author[M. O. Irfan]{Melis O. Irfan,$^{1,2}$\thanks{E-mail: mirfan@myuwc.ac.za}
\\ \\
$^{1}$Department of Physics and Astronomy, University of Western Cape, Cape Town 7535, South Africa\\
$^{2}$Department of Physics and Astronomy, Queen Mary University of London, London, E1 4NS, UK
\\
}

\date{Accepted XXX. Received YYY; in original form ZZZ}

\pubyear{2022}

\begin{document}
\label{firstpage}
\pagerange{\pageref{firstpage}--\pageref{lastpage}}
\maketitle

\begin{abstract}
We present a model for the full-sky diffuse Galactic synchrotron spectral index with an appropriate level of spatial structure for a resolution of 56 arcmin (to match the resolution of the Haslam 408\,MHz data). Observational data at 408\,MHz and 23\,GHz have been used to provide spectral indices at a resolution of 5 degrees. In this work we make use of convolutional neural networks to provide a realistic proxy for the higher resolution information, in place of the genuine structure. Our deep learning algorithm has been trained using 14.4\,arcmin observational data from the 1.4\,GHz Parkes radio continuum survey. We compare synchrotron emission maps constructed by extrapolating the Haslam data using various spectral index maps, of different angular resolution, with the Global Sky Model. We add these foreground maps to a total emission model for a 21\,cm intensity mapping experiment, then attempt to remove the foregrounds. The different models all display different spectral or spatial behaviour and so each provide a useful and different tool to the community for testing component separation techniques. We find that for an experiment operating using a cosine aperture taper beam with a primary Full Width at Half Maximum between 1.1 and 1.6 degrees, and the principal component analysis technique of foreground removal, there is a discernible difference between synchrotron spectral index models with a resolution larger than 5 degrees but that no greater resolution than 5 degrees is required. 
\end{abstract}

\begin{keywords}
Cosmology: diffuse radiation, Methods: statistical, Radio continuum: ISM
\end{keywords}



\section{Introduction}
\label{sec:intro}

Component separation has proven fundamental to observational cosmology; disentangling diffuse Galactic foregrounds from a cosmological signal of interest has been a central theme for Cosmic Microwave Background studies for decades \citep{ben03, gold09, gold11, leach08, psm, planck2014, planck16, planck18}. More recently a plethora of low-frequency ($ < 1.5$\,GHz) radio cosmology experiments have started observing with the aim of measuring the redshifted 21cm hydrogen line in order to probe Cosmic Dawn \citep{lwa, hera}, the Epoch of Reionisation \citep{paper, mwa} or the formation of large scale structure \citep{chime, bingo, fast, MKwhite}. At these frequencies diffuse Galactic synchrotron emission dwarfs the cosmological signal of interest; this emission is typically modelled as a power law with a  spectral index which scales the temperature across frequency. The synchrotron spectral index changes both spatially and gradually across frequency. Numerous works have investigated the mitigation of simulated foreground contamination on the detection of the simulated \hone \, signal, e.g. \citet{lauraFGsim, shawFGsim, bigotFGsim, alonsoFGsim, chapFGsim, intFGsim, mertFGsim, isaFGsim, bingoFGsim, steveFGsim, kpca, deepFGsim,  fastFGsim, paulaFGsim, skaFGsim}. It is vital that these simulated foregrounds contain an accurate level of spatial and spectral complexity to prevent a misleading simplification of the component separation problem. 

Publicly available repositories of diffuse Galactic foreground models are of enormous use to the community as they provide a test-bed on which to assess the qualities and deficiencies of component separation methods. Such resources include the Global Sky Model \citep{gsm}, which uses observational data to produce all-sky maps of diffuse Galactic emission between 10\,MHz and 5\,THz. The Global Sky Model (GSM) performs principal component analysis on empirical data sets to determine the statistically independent components within the sky maps and then interpolates this information to model the total diffuse emission temperature at any frequency within the 10\,MHz to 5\,THz range. Additionally, there are the Planck Sky \citep{psm} and Python Sky \citep{pysm} models which model the different physical components expected in the Galaxy due to differing emission mechanisms and use spectral index information to scale these maps across frequency.  

The 408 MHz all-sky map of \citet{haslam} is typically taken as a proxy for an all-sky map of synchrotron emission because synchrotron is thought to be the dominant diffuse emission at this frequency across the majority of the sky (excluding the central Galactic plane and a few specific molecular clouds). The simplest way to scale the Haslam map from 408\,MHz to any other frequency ($\nu$) is to use the power law parametrisation: 
\begin{equation}
T_{\nu}(p) = T_{408}(p) \left ( \frac{\nu}{408} \right)^{\beta},
\end{equation}
and assume a single value for the spectral index ($\beta$) which remains constant across pixels (p). However, the synchrotron spectral index is known to vary spatially due to energy losses of the relativistic, charged particles responsible for the emission \citep{ben03}. Previous works have highlighted the need to consider a variable synchrotron spectral index for the problem of 21\,cm foreground removal; demonstrating that the combination of a frequency-changing beam plus a spatially-varying spectral index is a more challenging foreground removal problem than just the situation of a frequency-changing beam plus a constant spectral index \citep{bern15, moz16, anstey}.

\citet{mamd} used the 408\,MHz all-sky map together with the {\it{WMAP}} 23\,GHz polarisation map to determine an all-sky model for the synchrotron spectral index between 0.408 and 23\,GHz. The 23\,GHz map of polarisation intensity was specifically used as at GHz frequencies synchrotron emission is no longer the dominant diffuse Galactic emission in intensity, but it is the only emission expected to be present in polarised intensity. To increase the signal-to-noise ratio of the polarised data the maps were smoothed to 5 degrees, enabling the production of an all-sky synchrotron spectral index map at 5 degree resolution. This map is used in both the Planck Sky Model and Python Sky Model to scale the 408\,MHz Haslam map across different frequencies. While it is generally accepted that the synchrotron spectral index changes spatially, the relationship between spectral index steepness and Galactic latitude is still not fully understood. Some spectral index maps, such as the 45 to 408\,MHz \citet{mu} map, show a range of values with the shallowest indices occurring within the Galactic plane. \citet{clive09} display six different spectral index maps from the literature, all with very different spatial features, some of which display a trend in spectral index steepness with Galactic latitude while some do not, and discuss how the visible spatial features depend on the variation assumptions made for the spectral index calculation. \citet{mu} attribute the presence of shallow spectral indices across the Galactic plane to free-free emission (spectral index $\sim -2.1$) due to the increased amount of warm ionised hydrogen within the Galactic plane. \citet{mamd} actually present three distinct maps of the synchrotron spectral index. Two are made from the Haslam 0.408 and {\it{WMAP}} 23\,GHz intensity maps; at 23 GHz in intensity several diffuse Galactic emissions are present across the sky. While model 1 models the free-free contribution at 23\,GHz, model 2 models both the free-free and anomalous microwave emission contributions. Both of these models also show shallower indices running across the plane. The third model, however, is made using the Haslam 408 MHz intensity map and the {\it{WMAP}} 23\,GHz polarised intensity map. As synchrotron emission is believed to be the only non-negligible emission present in polarised intensity at 23\,GHz, this is the only spectral index model that doesn't rely on a free-free or anomalous microwave emission model; this is also the only model out of the three which does not display shallow spectral indices within the Galactic plane. It is this spectral index model which is used in the Planck Sky Model and the Python Sky Model, therefore we follow suite and also use this spectral index map as our basis.

Previous works have already demonstrated the need for a spatially complex synchrotron spectral index model and provided such models; the highest resolution model available being at 5 degrees. We aim to determine if this level of spatial accuracy is enough or if a spectral index map with the same resolution as the Haslam data themselves would further complicate the foreground removal process. Inspired by the recent success of \citet{forse} who used Convolutional Neural Networks (CNNs) to learn high resolution features of thermal dust models from low resolution input models, we aim to simulate high resolution information for the \cite{mamd} spectral index map. We use a CNN trained on a spectral index map constructed using both Haslam and Parkes (CHIPASS) \citep{chipass} observational data. CNNs are a deep learning technique often employed for the task of image segmentation (assigning a label to each pixel of an image). We do not attempt a physically motivated model of synchrotron emission, as in \citet{magfield0} and \citet{magfield1} where the Galactic magnetic field itself is modelled. We simply attempt to construct a plausible model representative of the synchrotron spectral index for use in the testing of \hone \, data reduction pipelines. Previous models of the synchrotron spectral index and emission maps have used Gaussian realisations to provide additional spatial resolution; for instance \citet{destriped} use Gaussian realisations to provide a higher resolution estimate of the 408\,MHz map. Component separation methods, however, behave differently when attempting to clean Gaussian or non-Gaussian structure; \citet{skaFGsim} show that a variety of different techniques all struggle to deal with non-Gaussian foregrounds as viewed through a non-Gaussian (Airy) beam, while the same techniques do a better job of approaching the \hone \, power level when only Gaussian foregrounds are considered. Therefore we aim to determine if the additional complexity of non-Gaussian, 56 arcmin resolution spatial structure will provide a more accurately challenging test-bed for such techniques.

In this work we create a new all-sky spectral index template and assess how high resolution, non-Gaussian structure impacts the ability of a blind component separation method, Principal Component Analysis (PCA), to clean an emission cube of diffuse Galactic synchrotron emission in an attempt to measure the \hone \, auto-correlation power spectrum. Numerous foreground cleaning methods, both blind and parametric, are available for use and each method has different advantages and disadvantages when faced with spatial and spectral structure. It is not the aim of this work to provide a complete review of all existing component separation methods, as of such we choose to select one mainstream (i.e. often applied to intensity mapping data) technique to test here and make our spectral index map publicly available for the community to test alongside all other component separation techniques. We use our high resolution spectral index map, alongside the Haslam data at 408\,MHz to form diffuse synchrotron emission templates at various frequencies, which can then be compared to existing synchrotron emission models. In this work we compare four models for synchrotron emission: 1) the GSM model, 2) the Haslam data scaled using the 5 degree \citet{mamd} spectral index map, 3) the Haslam data scaled using a version of the 5 degree \citet{mamd} spectral index map, which has had higher resolution angular detail up to 56 arcmin added to it using Gaussian realisations and 4) the Haslam data scaled using the 5 degree \citet{mamd} spectral index map, which has had higher resolution angular detail up to 56 arcmin added to it using our trained CNN. 

We choose to only consider one foreground emission: diffuse Galactic synchrotron emission, so as to assess the effect our different models have on cleaning. In reality, low level diffuse free-free emission, extragalactic points sources, residual radio frequency interference (RFI) and possibly even residual ground emission pick-up may also be present in observational data. The simulations include \hone \, emission, white noise and diffuse synchrotron emission and each frequency channel is convolved with a frequency-dependant beam. We use the online \hone \, simulations repository \texttt{FastBox} \footnote{\url{https://github.com/philbull/FastBox}} to provide the test-bed set-up and so these simulations are specifically focused on foreground removal for a single-dish intensity mapping experiment using the MeerKAT dishes; an experiment such as MeerKLASS \citep{MKwhite}. However, the spectral index map that we have produced can be used by any experiment (single-dish or interferometric) alongside the Haslam data to get an estimate of diffuse synchrotron emission. 

The paper is laid out as follows: \autoref{sec:method} is split between the description of the construction of our high resolution spectral index map using CNNs and the description of our simulation test set-up i.e. each of the simulated components and the chosen component separation method. We also assess the success of our CNN in building a realistic level of high resolution structure within our spectral index map in \autoref{sec:method}. In \autoref{sec:res} we go on to use spherically-averaged auto-correlation power spectra to assess the impact of the different foreground maps on the cleaning ability of PCA. Our conclusions are presented in \autoref{sec:conc}.  

\section{Method}
\label{sec:method}

 \subsection{Constructing a high resolution spectral index map}
 
\begin{figure}
 \centering
  {\includegraphics[width=0.99\linewidth]{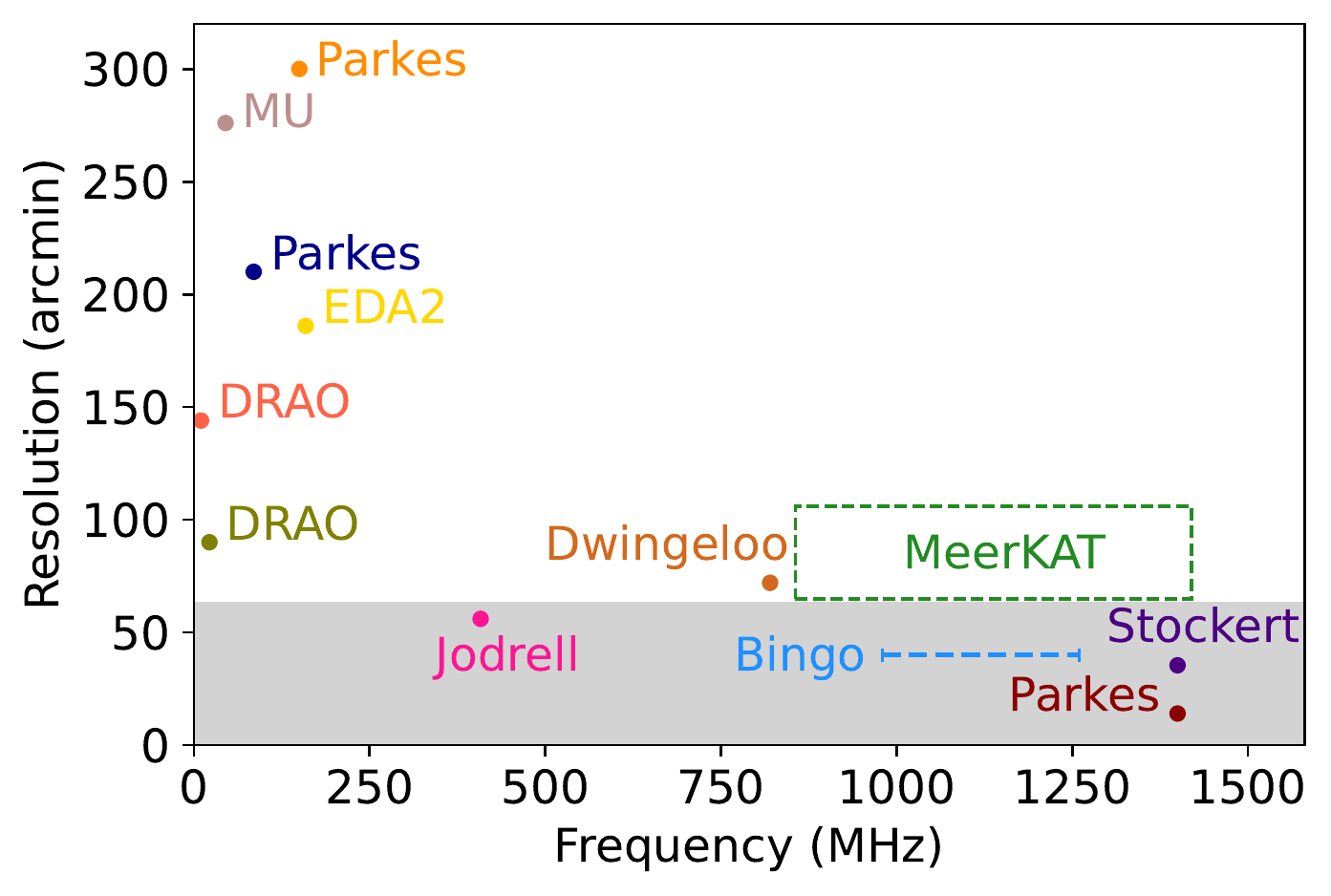}}\\
 \caption{Publicly available, full and partial sky observational data under 1.5\,GHz, as collated by the {\it{WMAP}} legacy archive. The survey central frequencies and resolutions are shown and each survey is labelled by observatory name; some surveys are made up from several telescope observations and in those cases only one observatory name has been listed. Full survey details can be found in the relevant papers: DRAO 10\,MHz \citep{drao}, DRAO 22\,MHz \citep{drao2}, MU \citep{mu}, Parkes 85\,MHz \citep{parkes}, Parkes 150\,MHz \citep{parkes}, EDA2 \citep{eda}, Jodrell \citep{haslam}, Dwingeloo \citep{dwing}, Parkes 1.4\,GHz \citep{chipass}, Stokert \citep{stock}. Two future radio surveys, using the Bingo and MeerKLASS telescopes, are also plotted using dotted lines. The grey strip highlights surveys with resolutions higher than or equal to 56 arcmin.}
 \label{fig:ldat}
   \end{figure}

Spectral index maps can be formed from data taken at two different frequencies ($\nu_{A}$ and $\nu_{B}$); the map produced gives the average spectral index per pixel required to scale the emission from $\nu_{A}$ to $\nu_{B}$ (as long as said emission can be modelled as a power law). The spectral index is an average over frequency if and only if it is believed that the spectral index changes across frequency. It is important to select two sets of observational data which can be believed to only contain the emission of interest. For example, in this work we want to create a map of the synchrotron spectral index and so we need two data sets observed across either frequencies or regions of the sky we believe to be dominated by synchrotron emission. The Haslam 408 MHz data is the standard proxy for diffuse synchrotron emission. To create a spectral index map using these data we required another synchrotron dominated observational data set of the same, or higher resolution. \autoref{fig:ldat} shows both full and partial radio continuum maps, publicly available from the {\it{WMAP}} legacy archive \footnote{\url{https://lambda.gsfc.nasa.govl}} as filled circles and two possible sources of future radio maps, from the MeerKLASS \citep{MKwhite} and Bingo \citep{bingo} \hone \, intensity mapping experiments as dotted lines. As the MeerKLASS data span a range of resolutions they are denoted using a dotted rectangle. The only two maps available with higher resolution than the Haslam data (which can also be seen on \autoref{fig:ldat}) are both at 1.4\,GHz. We choose to use the CHIPASS radio continuum survey, which is available at 14 arcmin resolution.

\begin{figure}
 \centering
  {\includegraphics[width=0.99\linewidth]{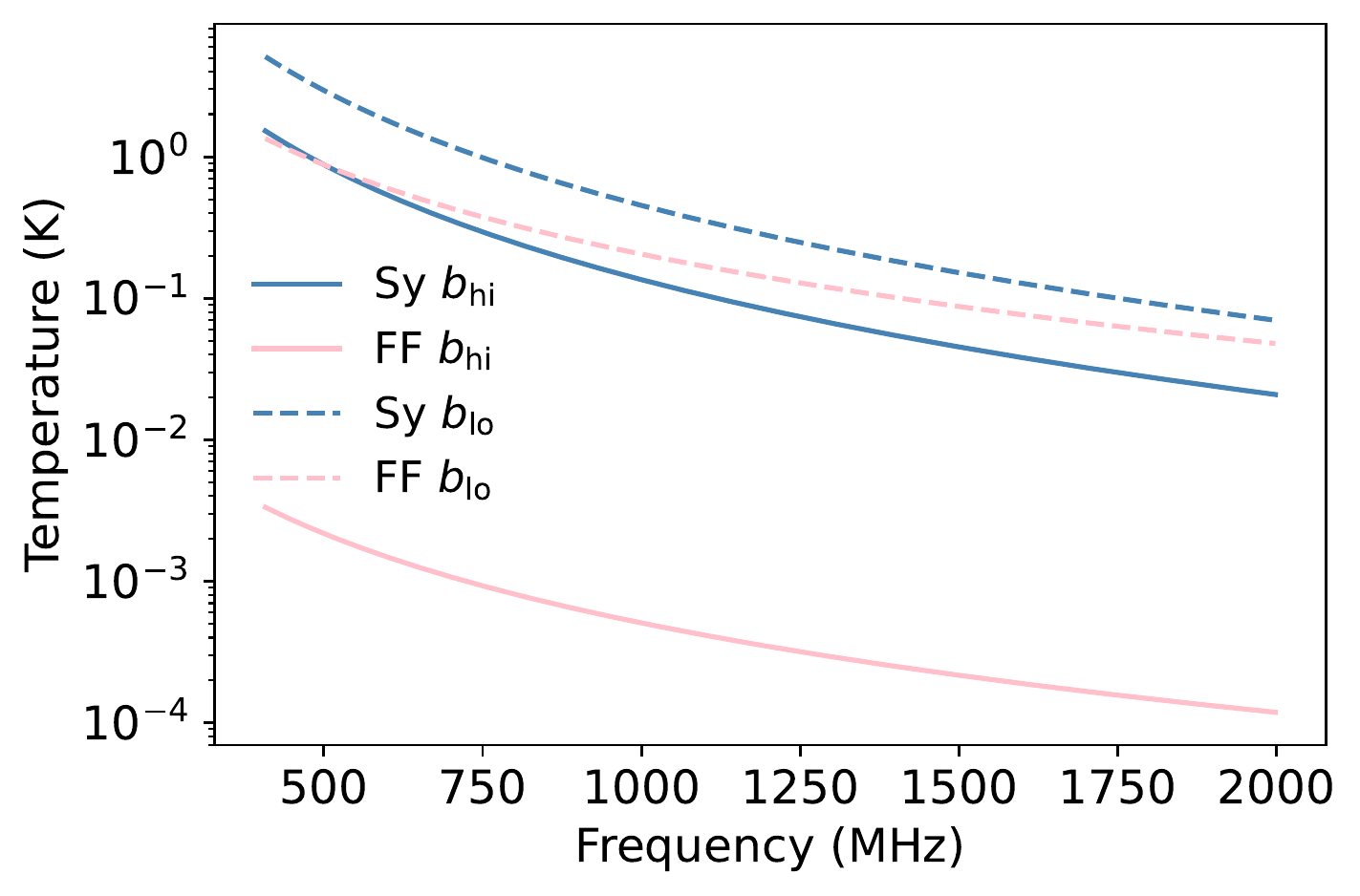}}\\
 \caption{Typical temperature variation across frequency for both synchrotron and free-free emission at high ($b_{{\rm{hi}}}$) and low ($b_{{\rm{lo}}}$) Galactic latitudes. The typical emission amplitude values were taken from the {\it{Planck}} full focal plane simulations and power law models were assumed for both with a spectral index of -2.7 for synchrotron and -2.1 for free-free emission.}
 \label{fig:coex}
   \end{figure}

We only consider publicly available maps under 1.5\,MHz, as at higher frequencies free-free emission is no longer a negligible component across the full sky. In \autoref{fig:coex} we plot the typical decrease in synchrotron and free-free emission temperatures across frequency for both low and high Galactic latitudes. The {\it{Planck}} full focal plane simulation data were used to provide the emission amplitudes \footnote{\url{https://pla.esac.esa.int/maps}}; we used the mean synchrotron and free-free emission temperatures within a 5 degree squared region centred at Galactic latitudes $75^{\circ}$ (to represent high latitudes) and $10^{\circ}$ (to represent low latitudes). Both emissions are modelled as a power law with a synchrotron spectral index of -2.7 \citep{beta} and a free-free spectral index of -2.1 \citep{ffbeta}.

 \subsubsection{The training and testing data}
\label{sec:tdata}

\begin{figure}
 \centering
  {\includegraphics[width=0.99\linewidth]{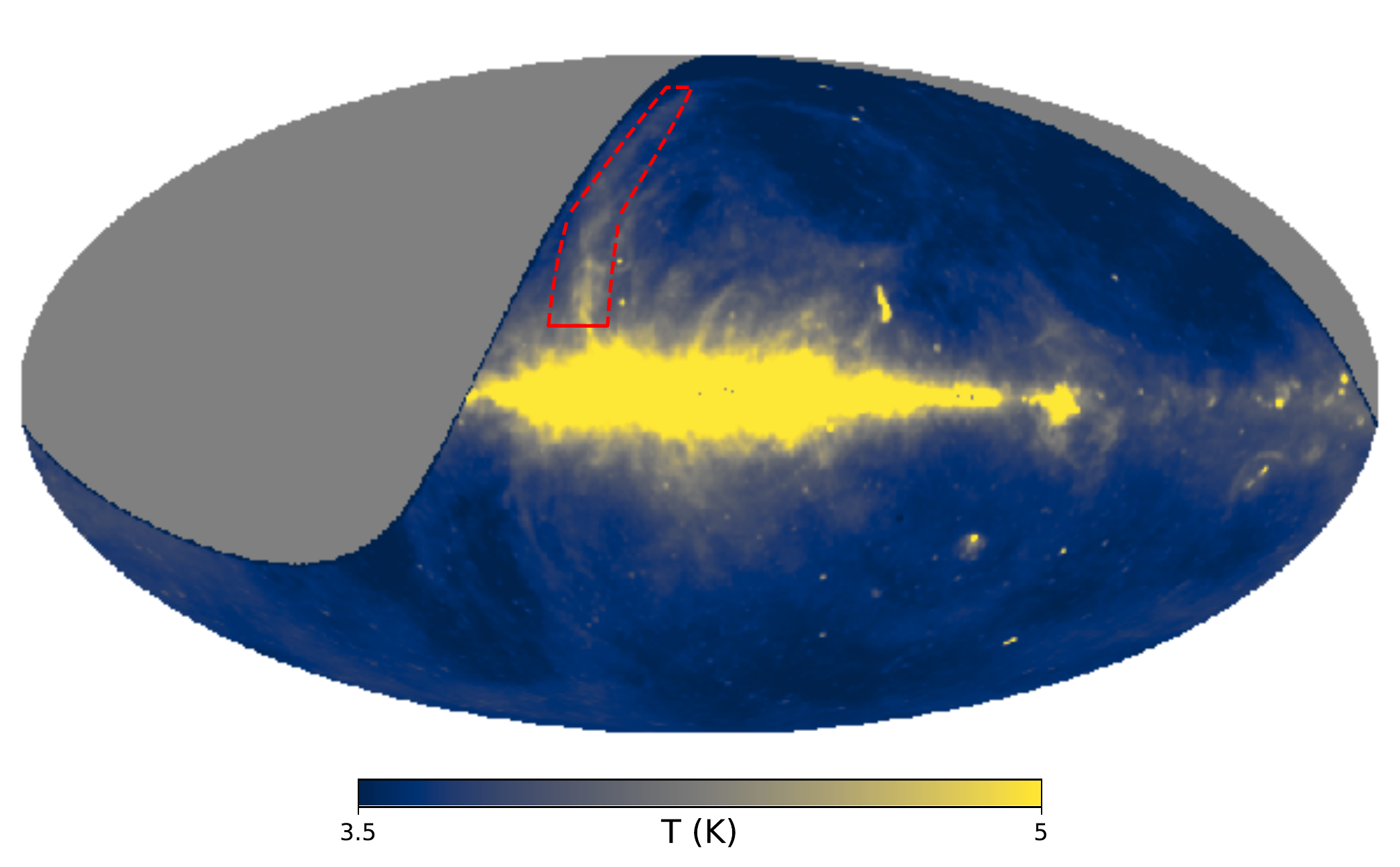}}\\
 \caption{The CHIPASS 1.4\,GHz map. The region used to create our synchrotron spectral index training data is highlighted in red.}
 \label{fig:cmap}
   \end{figure}

To train our network we used high resolution (14 arcmin) data from the CHIPASS experiment \citep{chipass} which cover the Southern sky at declinations < $25^{\circ}$. Even if the data were full sky, it would be risky to try and construct a full sky synchrotron spectral index between 0.408 and 1400\,MHz using the Haslam and CHIPASS data. At frequencies over 1\,GHz free-free emission is no longer believed to be negligible across the majority of the sky; free-free emission can contribute up to 50 per cent of the total emission close to the Galactic plane \citep{plat}. Therefore the full data were not used to train our network, instead we focused on the North Polar Spur (NPS) region which is believed to be a strong synchrotron emission feature \citep{nps}. This region is highlighted using red dotted lines in \autoref{fig:cmap} which shows the full CHIPASS data in Galactic coordinates. We restricted our training spectral index maps to come from this red dotted region only.

The 1.4\,GHz CHIPASS map was smoothed (assuming Gaussian beams) to 56 arcmin and used alongside the 56 arcmin, 408\,MHz Haslam map to produce a spectral index map:
 \begin{equation}
 \label{eq:the}
 \beta = \ln{\left (\frac{T_{\nu_{1}}}{T_{\nu_{0}}}\right ) } / \ln{ \left (\frac{\nu_{1}}{\nu_{0}} \right)},
 \end{equation}
 where $\nu_{0} = 408$\,MHz and $\nu_{1} = 1.4$\,GHz. The destriped, reprocessed version of the Haslam map \citep{destriped} was used. As this map is available at \texttt{HEALPix} \citep{healpix} ${\rm{N_{side}}}$ 512, we downgraded the ${\rm{N_{side}}}$ 1024 CHIPASS map to ${\rm{N_{side}}}$ 512. 
  
\begin{figure}
 \centering
  {\includegraphics[width=0.99\linewidth]{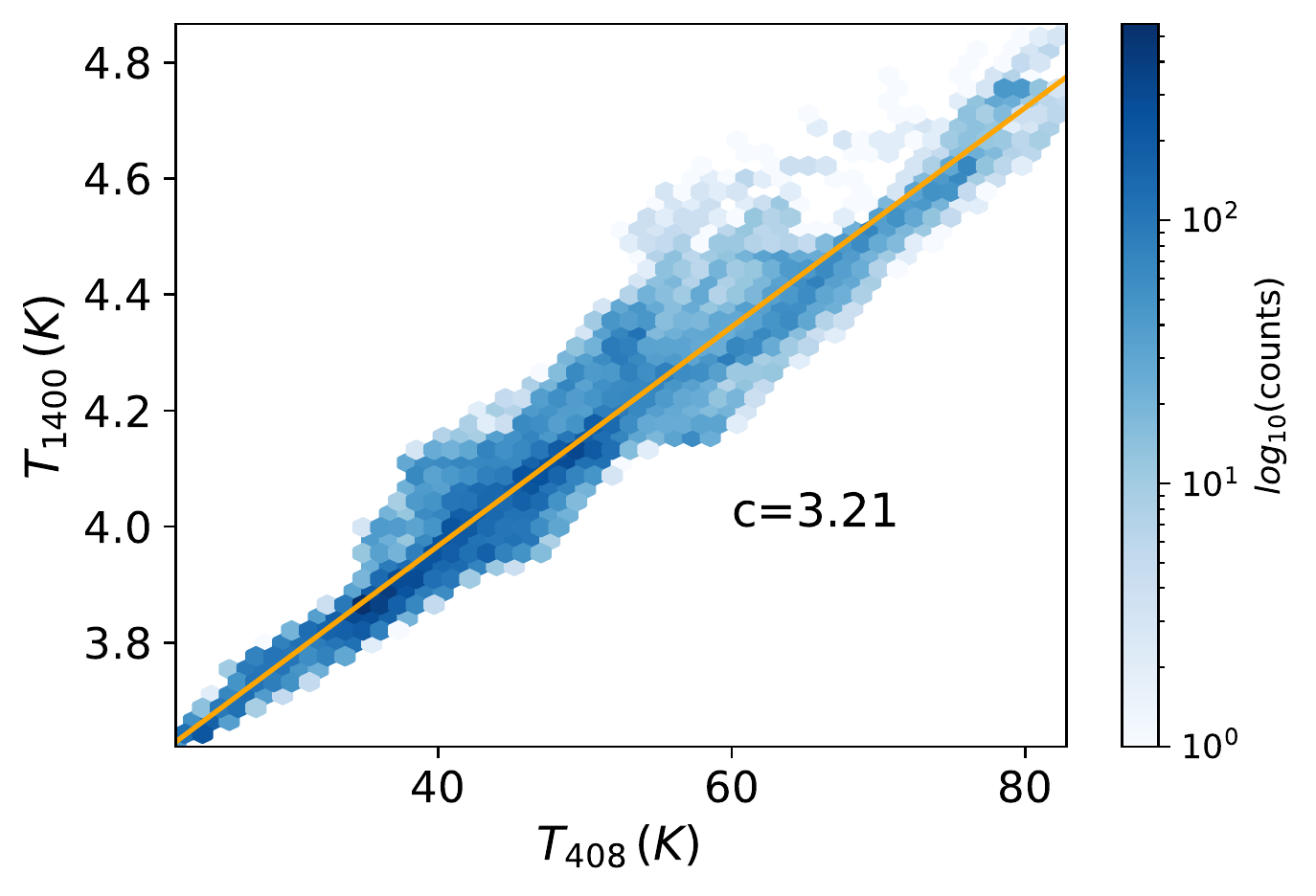}}\\
 \caption{Linear regression between the Haslam and CHIPASS data within the NPS region. The fitted offset is stated on the plot.}
 \label{fig:lreg}
   \end{figure}

For the synchrotron spectral index map to have the correct mean both the Haslam map and the CHIPASS data must have the correct zero-levels. As the Haslam data are so widely used as a proxy for synchrotron emission, considerable work has already been done to determine the map zero-level. \citet{wehus} used linear regression between multiple data sets to fit a zero-level of 8.9\,K to the Haslam data. We adopt that value and subtract it from the Haslam data. To find the zero-level of the CHIPASS data we then used the same temperature-temperature linear regression technique used in \citet{wehus}. The linear regression between the CHIPASS and Haslam data within our selected NPS region is given as:    
 \begin{equation}
T_{\rm{1400}}(p) = m \times T_{\rm{408}}(p)  + c,
\label{eq:weeq}
\end{equation}
where
 \begin{equation}
c = m \times c_{\rm 408} + c_{\rm 1400}, 
\end{equation}
where $ T_{\rm{1400}} $ is the temperature per pixel within the NPS region in the 1400\,MHz map,  $ T_{\rm{1400}} $ is the equivalent at 408\,MHz, $m$ is the gradient fitted from the linear regression and the fitted offset ($c$) is a combination of the offsets in both maps. By taking the Haslam offset ($ c_{\rm 408}$) as 8.9\,K, we could then calculate the CHIPASS offset. \autoref{fig:lreg} shows the linear regression within the NPS region between the Haslam and CHIPASS data; the Haslam data have already had the map zero-level of 8.9\,K removed. We found a fitted zero-level of 3.21\,K for the CHIPASS data; both the Haslam and CHIPASS maps were then used with their respective zero-levels subtracted to form a spectral index map. We show the spectral indices for our NPS region in  \autoref{fig:ainds}. In \autoref{appendix:A} we explore the uncertainty on our fitted zero-level of 3.21\,K by performing linear regression across regions of different Galactic latitude and examine how this uncertainty propagates to uncertainties within our final spectral index map. We find a 1$\sigma$ deviation of 0.1\,K on the zero-level, which effects both the mean level of the spectral indices determined, by 3 per cent, as well as their spatial variations, by 11 per cent at the resolution of 56 arcmin. 

 \begin{figure}
 \centering
  {\includegraphics[width=0.79\linewidth]{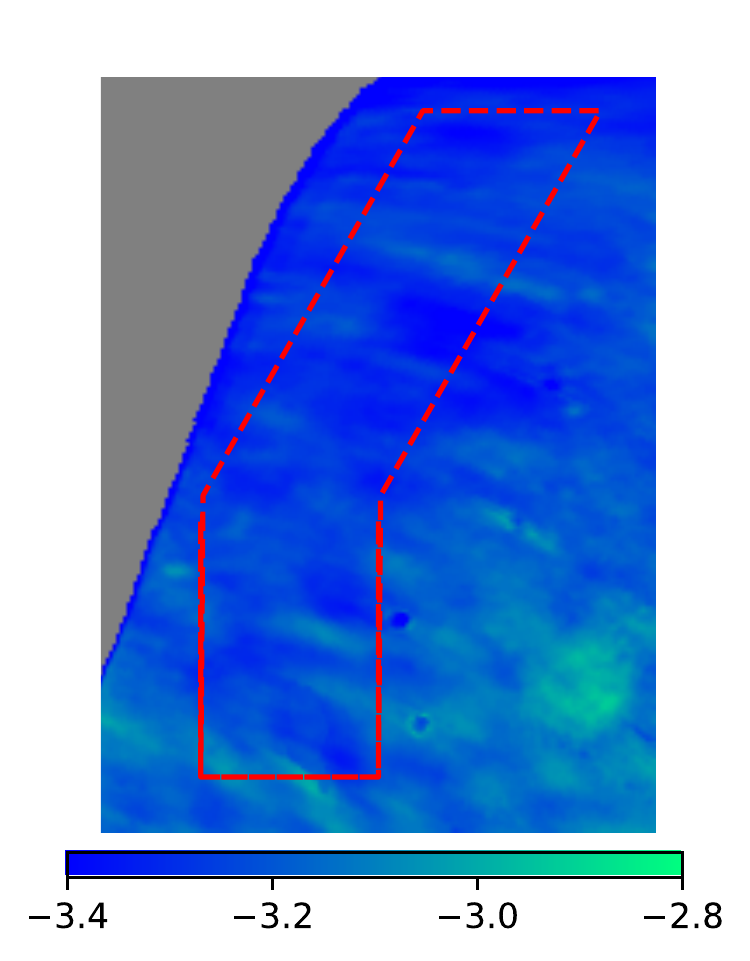}}\\
 \caption{Calculated synchrotron spectral indices between 408 and 1400\,MHz within the NPS region (highlighted using the red dotted line) at a resolution of 56 arcmin.}
 \label{fig:ainds}
   \end{figure}
 
The spectral index map produced was then smoothed to a resolution of 5 degrees, to provide both a low and high resolution perspective of the same data, and the data within our NPS region was cut up into numerous smaller sized spectral index maps. To increase the ease with which the network learnt the features across the numerous training spectral index maps, each separate map was normalised to contain spectral indices spanning from -1 to 1. Normalising the spectral index maps set the mean-level for all the maps to zero. This was not a problem however, as our interest is in the spatial variations around this mean; specifically the relationship between these variations at 56 arcmin and at 5 degree resolution. We selected 39 overlapping patches within this region, each spanning 7.3 degrees and having dimensions $64 \times 64$ pixels. A smaller number of patches of larger dimensions could have been chosen, but we opted for smaller images to reduce the number of features the network had to learn and to increase the number of training examples. Additionally, to maximise the number of training spectral index maps, we rotated each map three times by 90 degrees giving a total of $39 \times 4$ training maps. We formed these maps at both 56 arcmin and 5 degrees resulting in 156 pairs of maps.

 \subsubsection{The network}
\label{sec:tnet}  

CNNs can be built up using a variety of architectures. \citet{forse} use generative adversarial neural networks (GANs) to learn the 12 arcmin features from pairs of 12 and 80 arcmin thermal dust images taken from the GNILC all-sky thermal dust model \citep{gnilcDust}. Initially we began by training our CNN using a GAN architecture but found this structure failed to converge for our training data, possibly due to the small number of training information available as we could only use a small, synchrotron dominated fraction of the full sky. In the future, when experiments like MeerKLASS and Bingo publicly release large-area maps of the MHz sky, the GAN architecture can be revisited. For the current level of data availability however, we found the U-Net architecture optimal for our goals \citep{unet} and made use of the \texttt{Keras} \texttt{python} library. U-Net CNNs use a `U'-shaped symmetric structure of convolutional layers followed by deconvolution layers.

We have 156 pairs of synchrotron spectral index maps each of dimension $64 \times 64$ pixels. \autoref{tab:tabp} shows the details of each layer in the U-Net network used in this work. The first layer has 8 filters expanding the original map size from $64 \times 64$ to $64 \times 64 \times 8$, for the following layers the filters are doubled using a stride of 2. We picked a kernel size of 8, ensuring that the kernel size is divisible by the stride size to reduce the `checkerboard' effect in the final images \footnote{\url{https://distill.pub/2016/deconv-checkerboard/}}. Following the example of the generator \footnote{\url{https://github.com/ai4cmb/ForSE}} used in \citet{forse} we used the \texttt{LeakyReLU} activation function with a slope of 0.2, batch normalisation (to reinstate a zero mean and a variance of 1) after each convolution and a \texttt{tanh} activation for the final layer.

We held back 25 per cent of the spectral index map pairs for testing and used the other 75 per cent for training. To train our network we minimised the Mean Squared Error (MSE) loss function using the \texttt{Adam} optimiser for gradient descent with an initial learning rate of 0.0001. Use of the \texttt{ReduceLROnPlateau} option enabled the network to reduce the learning rate by a factor of 0.1 after seven iterations (epochs) with zero improvement in the loss function. The network was set to train in batches of size six and was allowed to stop whenever the loss function ceased to decrease after ten iterations. The testing spectral index maps were not used to train the network, but instead to evaluate it (see \autoref{sec:assess}).  

\begin{table}
\centering
\scalebox{0.9}{
\begin{tabular}{||c c c||} 
 \hline
 {\bf{Operation}} & {\bf{Dimensions}} & {\bf{Hyperparameters}} \\
 \hline\hline
 Input & 64 $\times$ 64 $\times$ 1  & \\
Convolution 2D & 64 $\times$ 64 $\times$ 8  & \\
Leaky ReLU &  & $\alpha = 0.2$\\
Batch normalisation &  & Momentum = 0.5\\
Convolution 2D & 32 $\times$ 32 $\times$ 16  & \\
Leaky ReLU &  & $\alpha = 0.2$\\
Batch normalisation & & Momentum = 0.5\\
Convolution 2D & 16 $\times$ 16 $\times$ 32  & \\
Leaky ReLU & & $\alpha = 0.2$\\
Batch normalisation & & Momentum = 0.5\\
Convolution 2D & 8 $\times$ 8 $\times$ 64  & \\
Leaky ReLU & & $\alpha = 0.2$\\
Batch normalisation & & Momentum = 0.5\\
Convolution 2D & 4 $\times$ 4 $\times$ 128  & \\
Leaky ReLU &  & $\alpha = 0.2$\\
Batch normalisation &  & Momentum = 0.5\\
Convolution 2D & 2 $\times$ 2 $\times$ 256  & \\
Leaky ReLU &  & $\alpha = 0.2$\\
Batch normalisation & & Momentum = 0.5\\
Up sampling 2D & & \\
Convolution 2D & 4 $\times$ 4 $\times$ 128  & \\
Leaky ReLU & & $\alpha = 0.2$\\
Batch normalisation &  & Momentum = 0.5\\
Up sampling 2D & & \\
Convolution 2D & 8 $\times$ 8 $\times$ 64  & \\
Leaky ReLU &  & $\alpha = 0.2$\\
Batch normalisation & & Momentum = 0.5\\
Up sampling 2D & & \\
Convolution 2D & 16 $\times$ 16 $\times$ 32  & \\
Leaky ReLU & & $\alpha = 0.2$\\
Batch normalisation &  & Momentum = 0.5\\
Up sampling 2D & & \\
Convolution 2D & 32 $\times$ 32 $\times$ 16  & \\
Leaky ReLU &  & $\alpha = 0.2$\\
Batch normalisation &   & Momentum = 0.5\\
Up sampling 2D & & \\
Convolution 2D & 64 $\times$ 64 $\times$ 8  & \\
Leaky ReLU &  & $\alpha = 0.2$\\
Batch normalisation & & Momentum = 0.5\\
Up sampling 2D & & \\
Convolution 2D & 64 $\times$ 64 $\times$ 1 & \\
Tanh &  & \\
 \hline \hline
\end{tabular}}
\caption{The U-Net architecture used in this work.}
\label{tab:tabp}
\end{table}

 \subsubsection{Processing full-sky data}
\label{sec:fdata}    
 After the network had been trained the goal was to take the full-sky spectral index map of \citet{mamd} at 5 degree resolution and use it as the input for the network to generate a 56 arcmin resolution version. We obtained the \citet{mamd} map from the {\it{Planck}} Full Focal Plane simulations \citep{ffp}, using \autoref{eq:the} where $T_{\nu_{0}/\nu_{1}}$ are the simulated synchrotron temperature maps at 353 and 217\,GHz. The {\it{Planck}} Full Focal Plane synchrotron simulations are available at ${\rm{N_{side}}}$ 2048, so we downgraded the synchrotron spectral index map to ${\rm{N_{side}}}$ 512. The $12 \times 512 \times 512$ pixels were then projected into 768 spectral index maps of dimensions $64 \times 64$ pixels using \texttt{healpy} functions \citep{healpy}.
 
Once the 768 generated spectral index maps were projected back onto the sphere, we enlisted the technique of Cycle-spinning \citep{cycle} to remove any border effects caused as a result of the projections between the 3D sphere and the 2D image plane. Cycle-spinning involves rotating the full-sky, 5 degree spectral index map, breaking up the map into 2D maps which are then used by the network to estimate high resolution 2D maps, re-projecting these maps to form a high resolution all-sky map and then performing the inverse rotation. We perform 12 rotations in total, plus the original map: Axis (X,Y): $(0^{\circ}, 0^{\circ})$, $(45^{\circ}, 0^{\circ})$, $(0^{\circ}, 90^{\circ})$, $(-45^{\circ}, 0^{\circ})$, $(0^{\circ}, -90^{\circ})$, $(45^{\circ}, 90^{\circ})$, $(45^{\circ}, -90^{\circ})$, $(-45^{\circ}, 90^{\circ})$, $(-45^{\circ}, -90^{\circ})$, $(90^{\circ}, 90^{\circ})$, $(90^{\circ}, -90^{\circ})$, $(-90^{\circ}, 90^{\circ})$ and $(-90^{\circ}, -90^{\circ})$. These thirteen maps were then averaged together. Lastly we smoothed the maps from the pixel resolution (6.87 arcmin for an ${\rm{N_{side}}}$ 512 map) to 56 arcmin. 

\begin{figure}
 \centering
  {\includegraphics[width=0.99\linewidth]{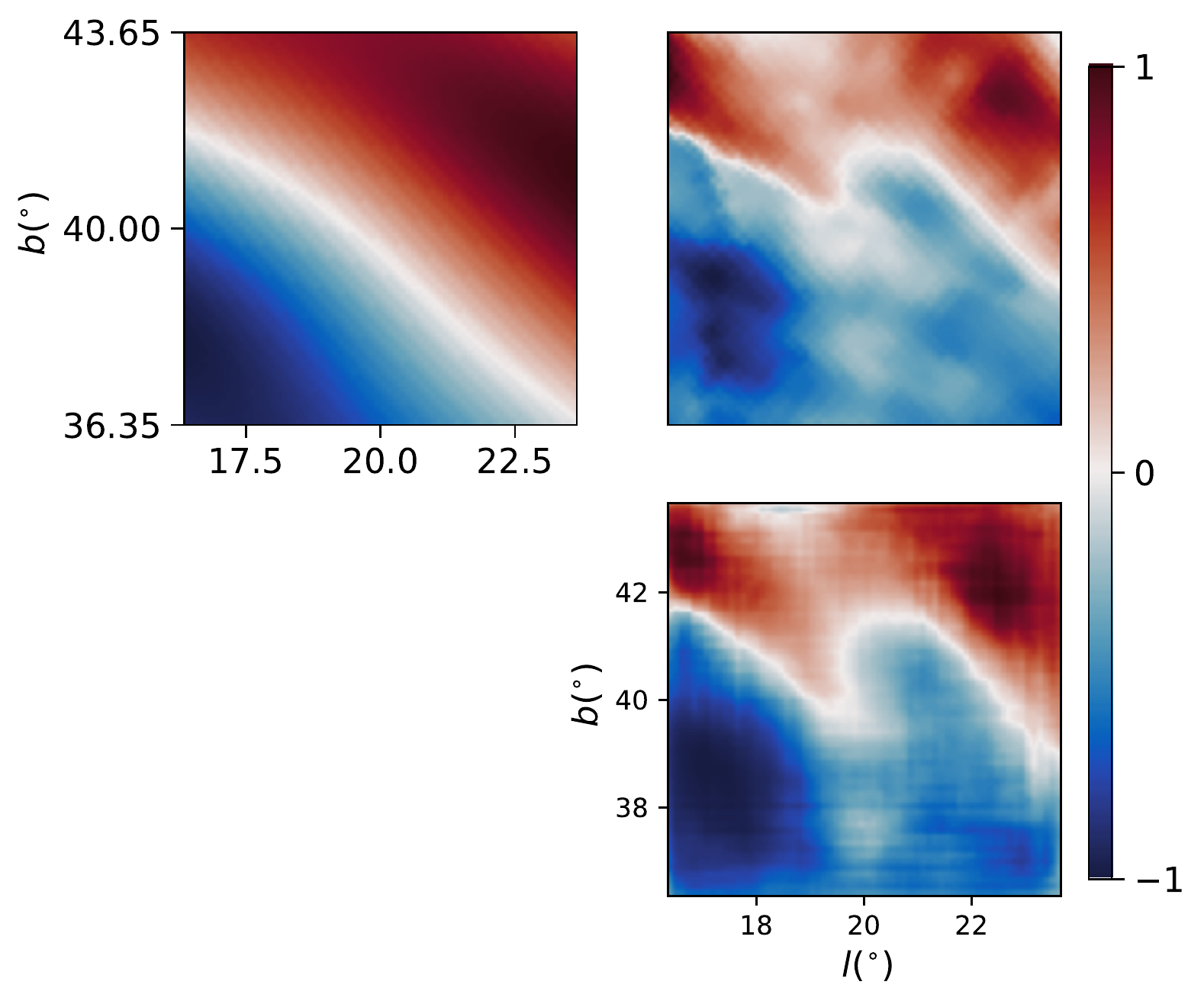}}\\
 \caption{{\it{Top left:}} One of the normalised 5 degree spectral index training maps, {\it{top right:}} the equivalent 56 arcmin version of the map, {\it{bottom:}} the equivalent map generated by the trained network.}
 \label{fig:tmaps}
   \end{figure}

 \subsubsection{Assessing the synchrotron spectral index map}
\label{sec:assess}    

An example of a training spectral index map pair given to the network is shown in \autoref{fig:tmaps}. The top left panel shows a normalised $64 \times 64$ pixel spectral index map at 5 degree resolution, while the top right shows the same but for the 56 arcmin spectral index map. In the lower panel we have the map generated by the trained network having been given the 5 degree map as input. The network can be seen to have learnt the key features in the high resolution map. There are clearly pixel effects in the generated map, such as a faint checkerboard effect and the odd spurious (not true to the high resolution map) pixel. However, as we are not attempting to create a spectral index map at the pixel resolution (6.87 arcmin for an ${\rm{N_{side}}}$ 512 map) these spurious effects get removed by the Cycle spinning and smoothing processes described in \autoref{sec:fdata}.  
   
   \begin{figure}
 \centering
  {\includegraphics[width=0.99\linewidth]{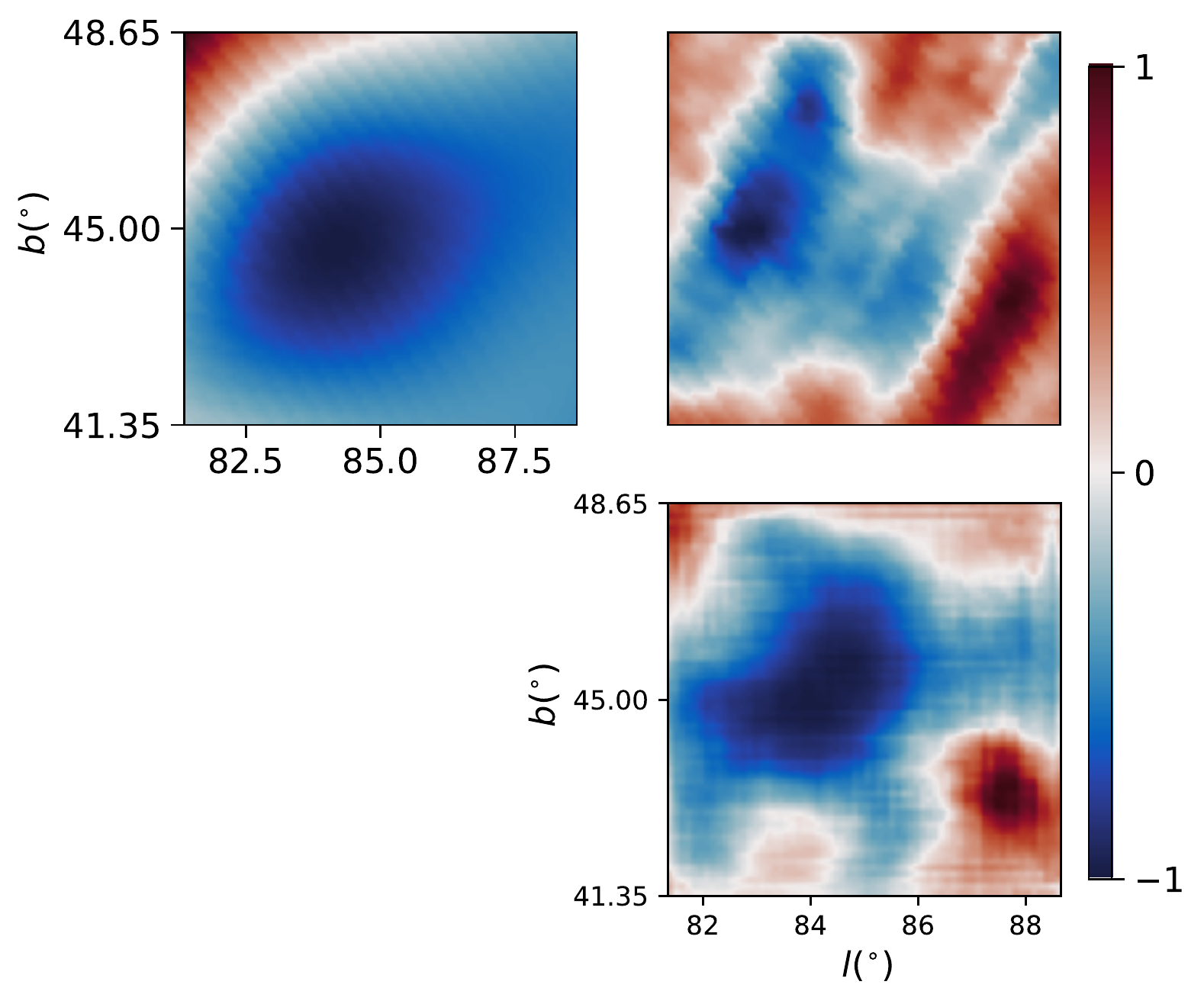}}\\
 \caption{{\it{Top left:}} One of the normalised 5 degree spectral index test maps, {\it{top right:}} the equivalent 56 arcmin version of the map, {\it{bottom:}} the equivalent map generated by the trained network.}
 \label{fig:ttmaps}
   \end{figure}
   
\begin{figure}
 \centering
  {\includegraphics[width=0.9\linewidth]{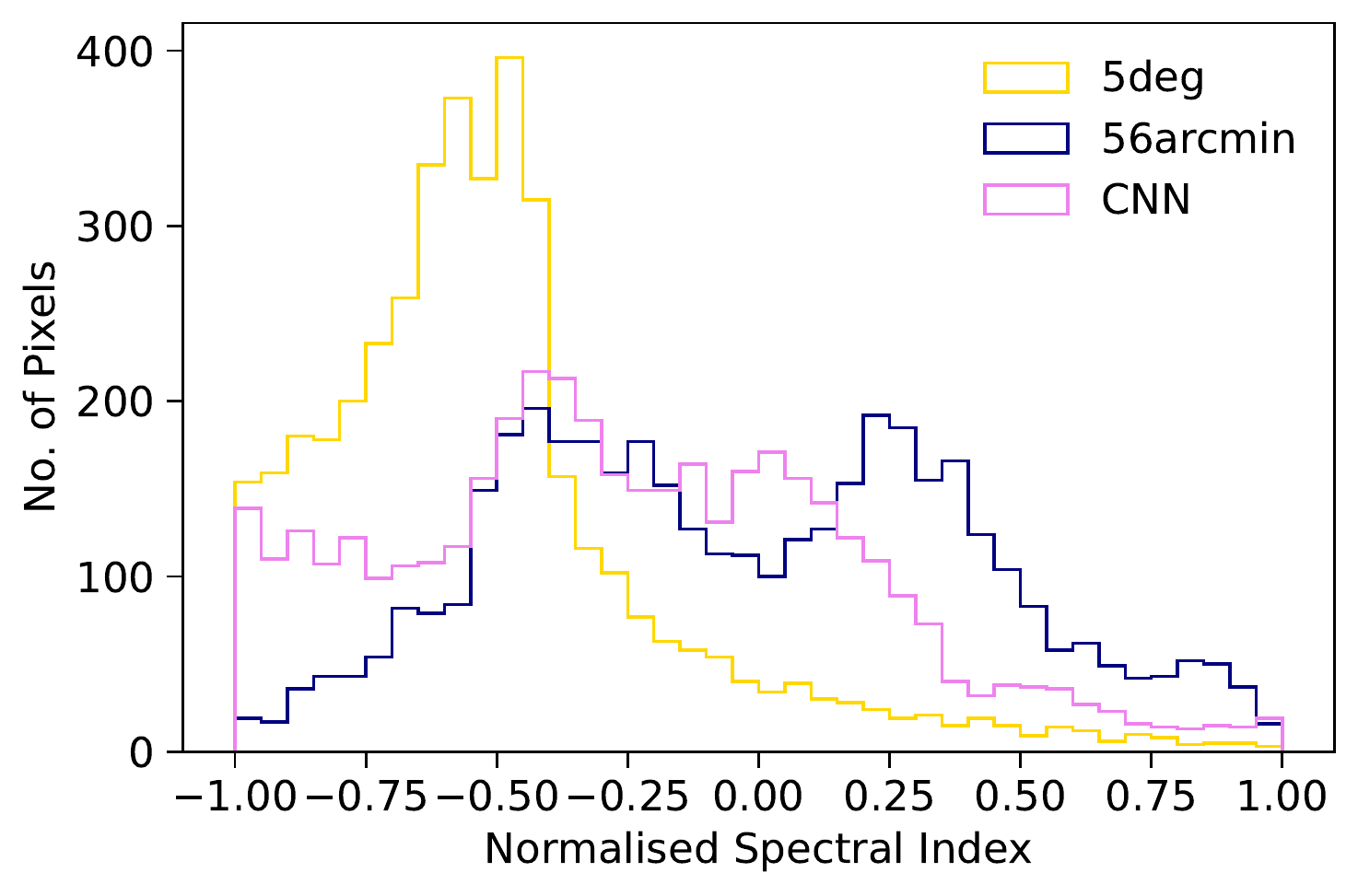}}\\
 \caption{Histogram distribution of the normalised spectral indices in a test high and low resolution maps and the network generated high resolution model. These distributions are for the same data show in \autoref{fig:ttmaps}}
 \label{fig:histp}
   \end{figure}

To see how well the network can generate high resolution spectral index maps from a low resolution map not used in the training we use a map from the test subset. The top left panel of \autoref{fig:ttmaps} shows a low resolution map from the test subset, and the accompanying high resolution map in the top right panel. The network generated map is shown in the lower panel. It can be seen that the reproduction of the small scale structure is far less faithful to the true high resolution map than in the case of the training data shown in \autoref{fig:tmaps}. However, on visual inspection the level of spatial structure in the generated map seems appropriately detailed. The histogram distributions of map pixels can be used as a method to assess image complexity. In \autoref{fig:histp} we show the histograms for the same maps displayed in \autoref{fig:ttmaps}. The low resolution training map has one strong peak but other than that has a very flat histogram distribution, while both the 56 arcmin map and the network generated map show considerably more structure. The generated test maps are not correct, in that they are not identical to the true 56 arcmin test data, but they contain high resolution spatial structure and they remain faithful to the large scale structure in the map therefore they can be used to provide a high resolution spectral index map.     

   \begin{figure}
 \centering
  {\includegraphics[width=0.9\linewidth]{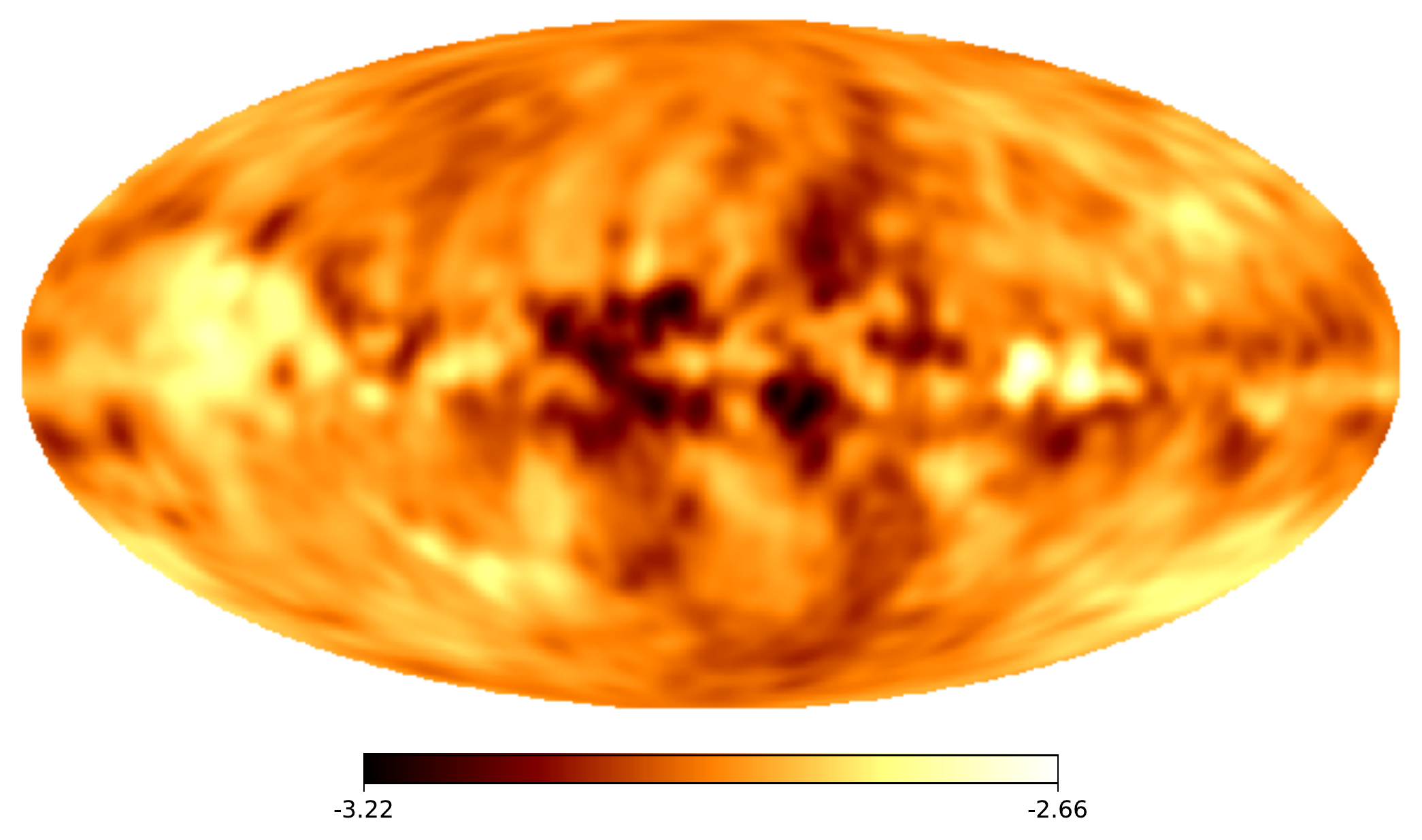}}\\
    {\includegraphics[width=0.9\linewidth]{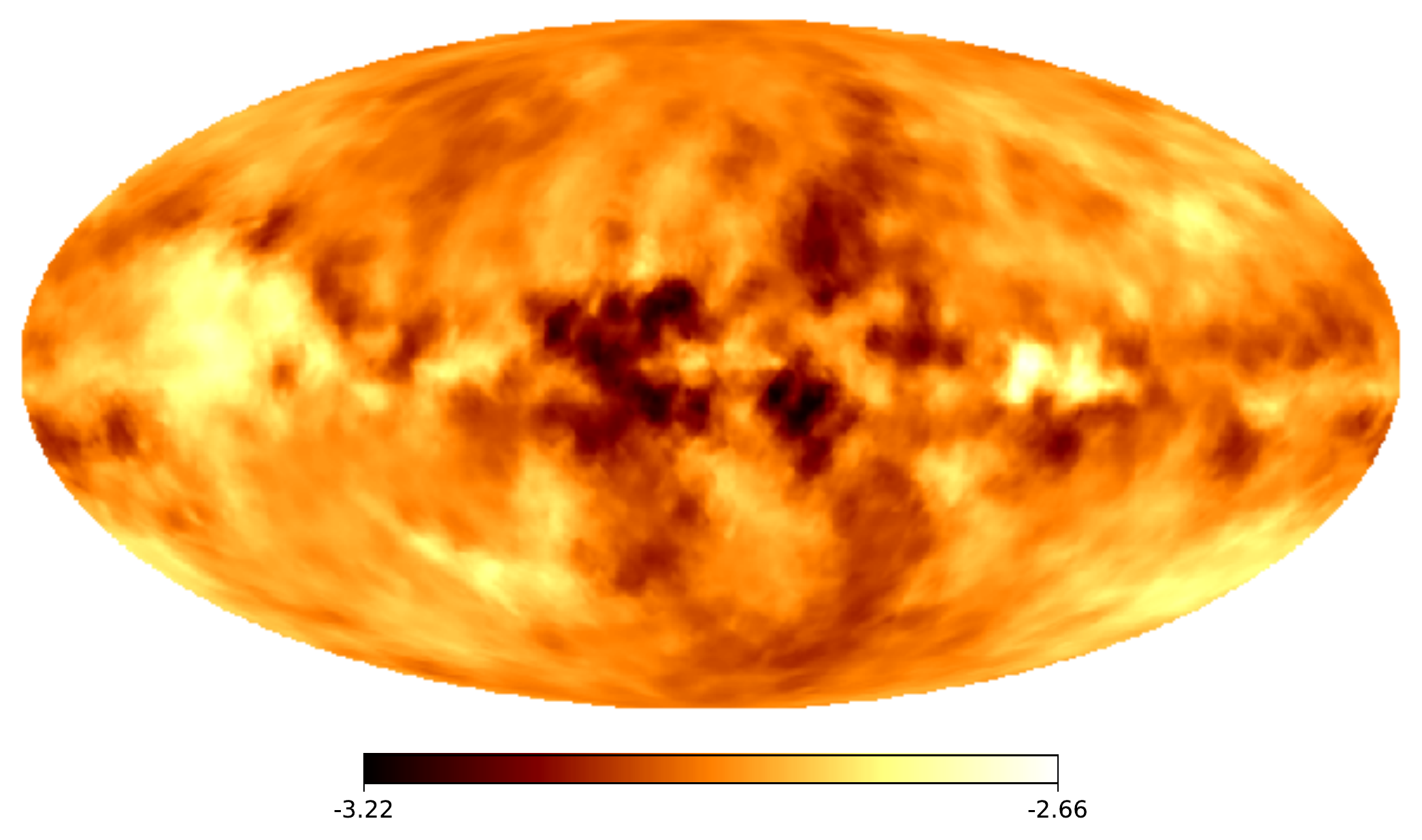}}\\
 \caption{ Full sky synchrotron spectral index between 408\,MHz and 23\,GHz at 5 degrees ({\it{top}}) and 56 arcmin ({\it{bottom}}) resolution.}
 \label{fig:allmaps}
   \end{figure} 
   
   \begin{figure}
 \centering
  {\includegraphics[width=0.99\linewidth]{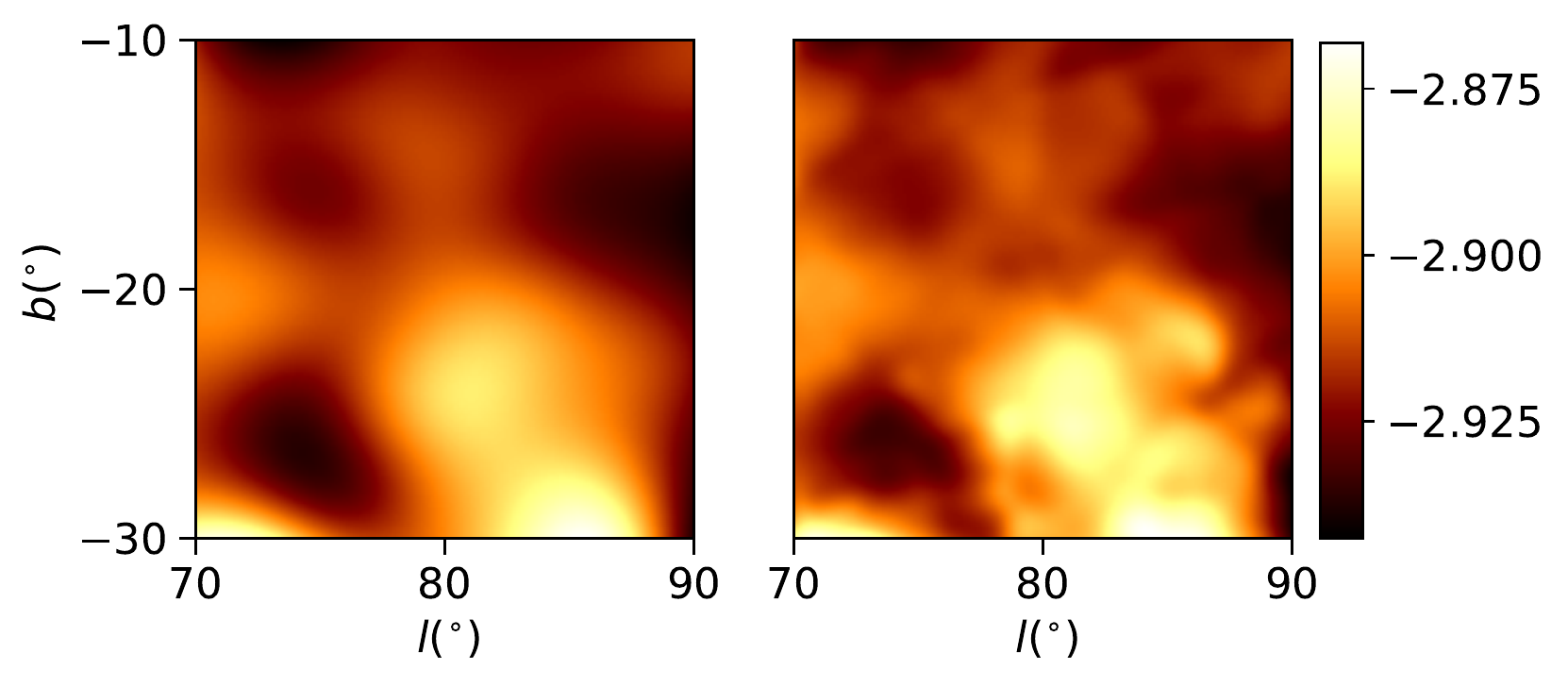}}\\
 \caption{A 20$^{\circ}$ by 20$^{\circ}$ zoom-in of the all-sky spectral index map at ({\it{left:}}) 5 degree and ({\it{right:}}) 56 arcmin resolution.}
 \label{fig:zmaps}
   \end{figure} 
   
\begin{figure}
 \centering
  {\includegraphics[width=0.99\linewidth]{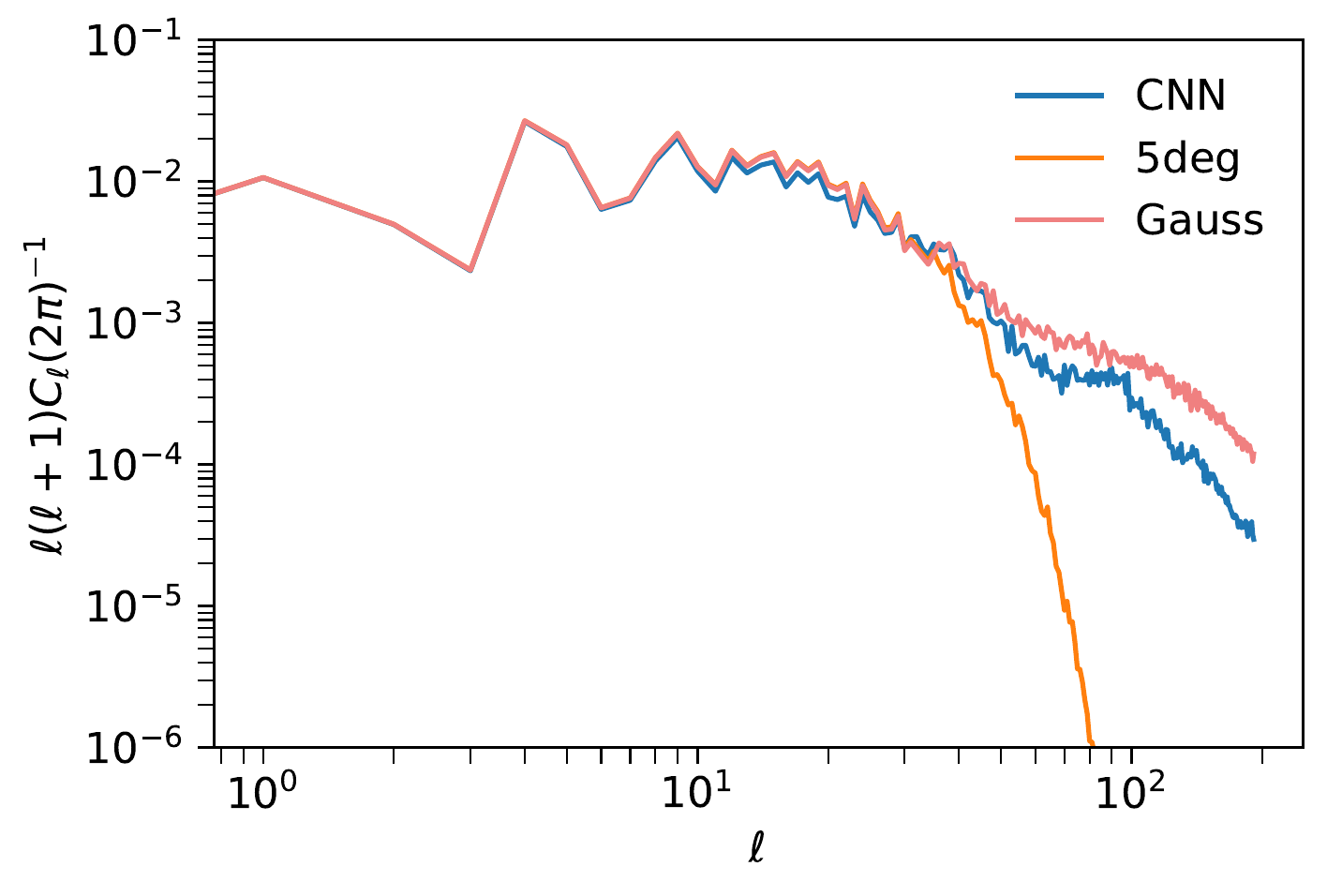}}\\
 \caption{Angular power spectra of the 5 degree, CNN and Gaussian generated synchrotron spectral index map.}
 \label{fig:cls}
   \end{figure}

The original 5 degree resolution, all-sky spectral index map of \citet{mamd} is displayed in the top panel of \autoref{fig:allmaps}, whilst our 56 arcmin version is presented in the lower panel. Our 56 arcmin map is publicly available \footnote{\url{https://github.com/melisirfan/synchrotron_emission}}. \autoref{fig:zmaps} selects a smaller region of the all-sky maps to clearly demonstrate the additional detail in the high resolution version of the \citet{mamd} spectral index map. The \texttt{healpy} \texttt{anafast} library was used to calculate the angular power spectra for both the original and our new spectral index map, shown in \autoref{fig:cls}. The angular power spectrum for the original map drops off at $\ell \sim 40$ / $\theta \sim 4.5^{\circ}$ while the CNN generated spectral map power only starts to drop off at $\ell \sim 150$ / $\theta \sim 1.2^{\circ}$. We can see that the high resolution spectral index map suffers from a slight power loss (between 5 and 15 percent) from the original map between $\sim$ 6 and 18 degrees, as our CNN fails to perfectly reconstruct high-resolution images different to those that it has been trained with. We also add the angular power spectrum of a synchrotron spectral index map generated by adding high resolution structure to the 5 degree map using a Gaussian realisation with Gaussian structure up to 56 arcmin. The high resolution Gaussian structure was added using a power-law in $\ell$: $A \times (36/\ell)^{\beta}$, where the amplitude was set by the 5 degree spectral index map power at $\ell = 36$ and $\beta = 2.4$ following the parametrisation for synchrotron emission detailed in \cite{santos05}. This will be pertinent for the next section, where we discuss the full simulation and four options for synchrotron emission models.   

 \subsection{The simulation}
 \label{sec:theset}    
 
Having constructed a high resolution spectral index map, we wished to investigate whether or not such a map would impact the results of a foreground cleaning simulation. We used the \hone \, simulation suite \texttt{FastBox} to set up two data cubes of dimensions $128\times128\times128$. Both cubes cover a 3600 square degree region with a pixel resolution of 0.47$^{\circ}$  (60$^{\circ}$ across 128 pixels). The first data cube spans the frequency range 1220 to 1363\,MHz over 128 pixels, giving a frequency resolution of 1.1\,MHz and the seconds cube spans from 1084 to 1226\,MHz, also with a frequency resolution of 1.1\,MHz. The total data cubes consist of \hone \, emission plus diffuse synchrotron emission and are convolved with the MeerKAT beam, instrumental white noise is then added into the simulations after the beam convolution. 

The \texttt{FastBox} \hone \, signal is simulated as:
\begin{equation}
\Delta T_b(\vec{x}, z) = \overline{T}_b(z)\, b_{\rm HI}(z)\, \delta_m(\vec{x}, z),
\end{equation}
details of the mean brightness temperature, HI bias and fractional HI density used can be found in \cite{kpca}. Log-normal transformations of the Gaussian field are applied to ensure a physical density distribution and the effect of Redshift Space Distortions are included by shifting each 3D pixel of the transformed field. 

\begin{figure}
 \centering
  {\includegraphics[width=0.69\linewidth]{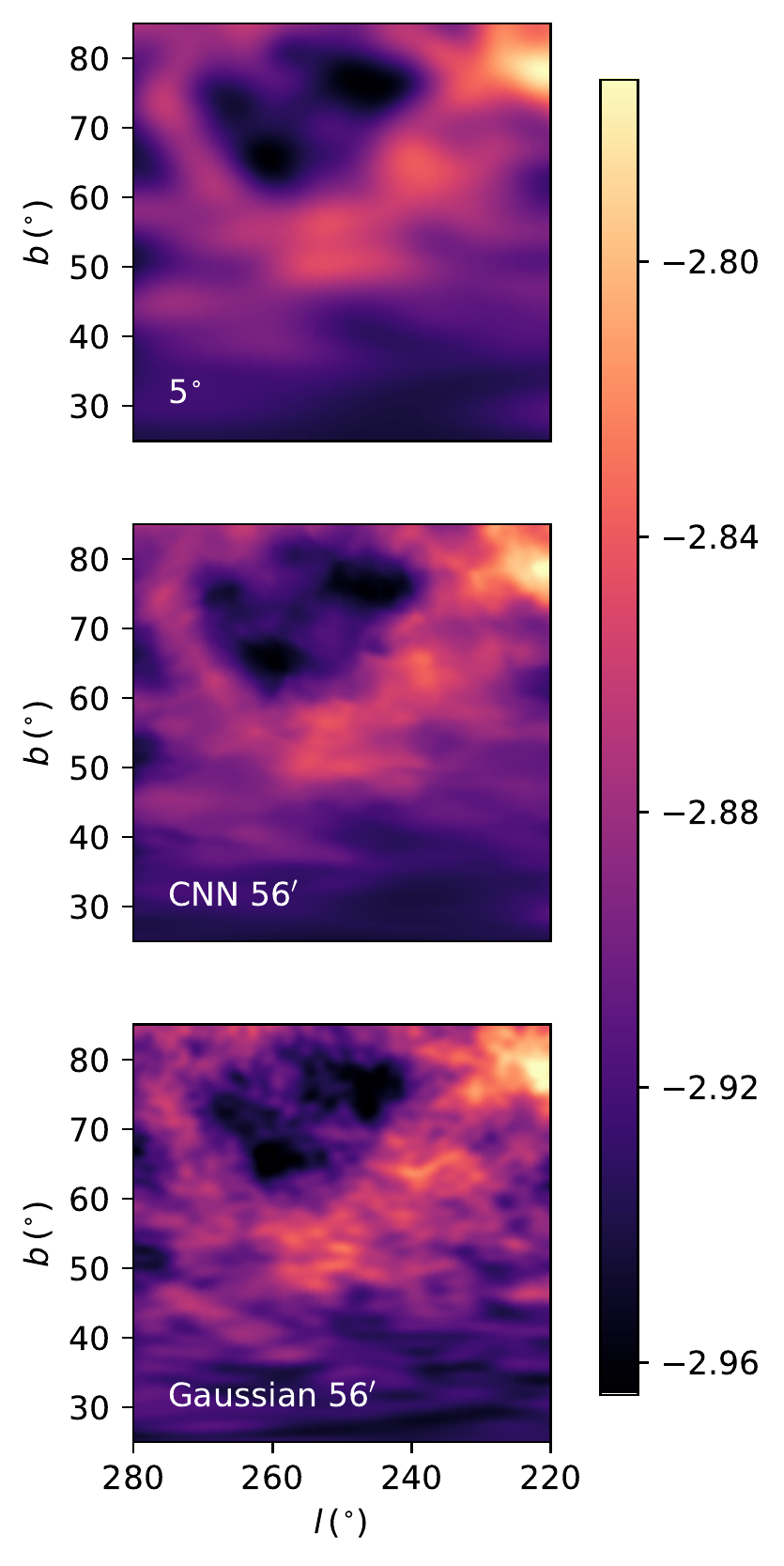}}\\
 \caption{The three spectral index maps investigated in this work: the 5$^{\circ}$ model is on top, the CNN model is in the middle and the model with high resolution information filled in using a Gaussian realisation is on the bottom row.}
 \label{fig:bmaps}
   \end{figure}

For the foreground contribution we only simulate diffuse synchrotron emission, as we wish to explore different synchrotron models. All our foreground models cover the 3600 degree square region of $25^{\circ} <$ Galactic latitude ($b$)\,$< 85^{\circ}$ and $220^{\circ} <$ Galactic longitude ($l$)$\,< 280^{\circ}$. The first model is the Global Sky Model; the GSM is in fact a model of the total diffuse Galactic emission at the user-selected frequency. As the simulated data are at high Galactic latitude and MHz frequencies the assumption is that synchrotron emission will be the dominant emission, but this assumption may not necessarily hold true within the GSM map which will contain some fractional level of extrapolated free-free, anomalous microwave and thermal dust emission. Models two, three and four are all models of pure synchrotron emission assuming a power law model form and using the Haslam 56 arcmin data as the emission amplitude at 408\,MHz. These three models only differ in the resolution of the spectral index map used. The 5 degree spectral index map of \citet{mamd} is the base for all three spectral index maps and provides all the spectral index information for model two. Model three uses the new spectral index map produced by this work, which has high resolution information between 56 arcmin and 5 degrees provided by our CNN and model four has the high resolution information simulated using a Gaussian realisation. \autoref{fig:bmaps} shows the three different spectral indices per pixel for our 3600 square degree region. Common large-scale structure is clear in all maps but the 5 degree map is clearly missing the high resolution structure seen in the other two. The Gaussian high resolution map looks to be simply more noisy than the CNN map, as opposed to containing high resolution structure. The four foreground emission models are summarised in \autoref{fgmod}.   

\begin{table}
\centering
\begin{tabular}{||c c c||} 
 \hline
 {\bf{Model name}} & {\bf{Emission}} & {\bf{Resolution}} \\
 \hline\hline
GSM & total diffuse Galactic & 56 arcmin  \\
5$^{\circ}$ & diffuse Galactic synchrotron & 56 arcmin/ 5$^{\circ}$ \\
CNN & diffuse Galactic synchrotron & 56 arcmin/ 56 arcmin \\
Gaussian & diffuse Galactic synchrotron & 56 arcmin/ 56 arcmin \\
 \hline \hline
\end{tabular}
\caption{The four foreground emission models investigated in this work and their resolutions. The last three models require both amplitude and spectral index templates which is why they have two resolutions stated for each; the first for the amplitude, the second for the spectral index.}
\label{fgmod}
\end{table}

\begin{figure}
 \centering
  {\includegraphics[width=0.99\linewidth]{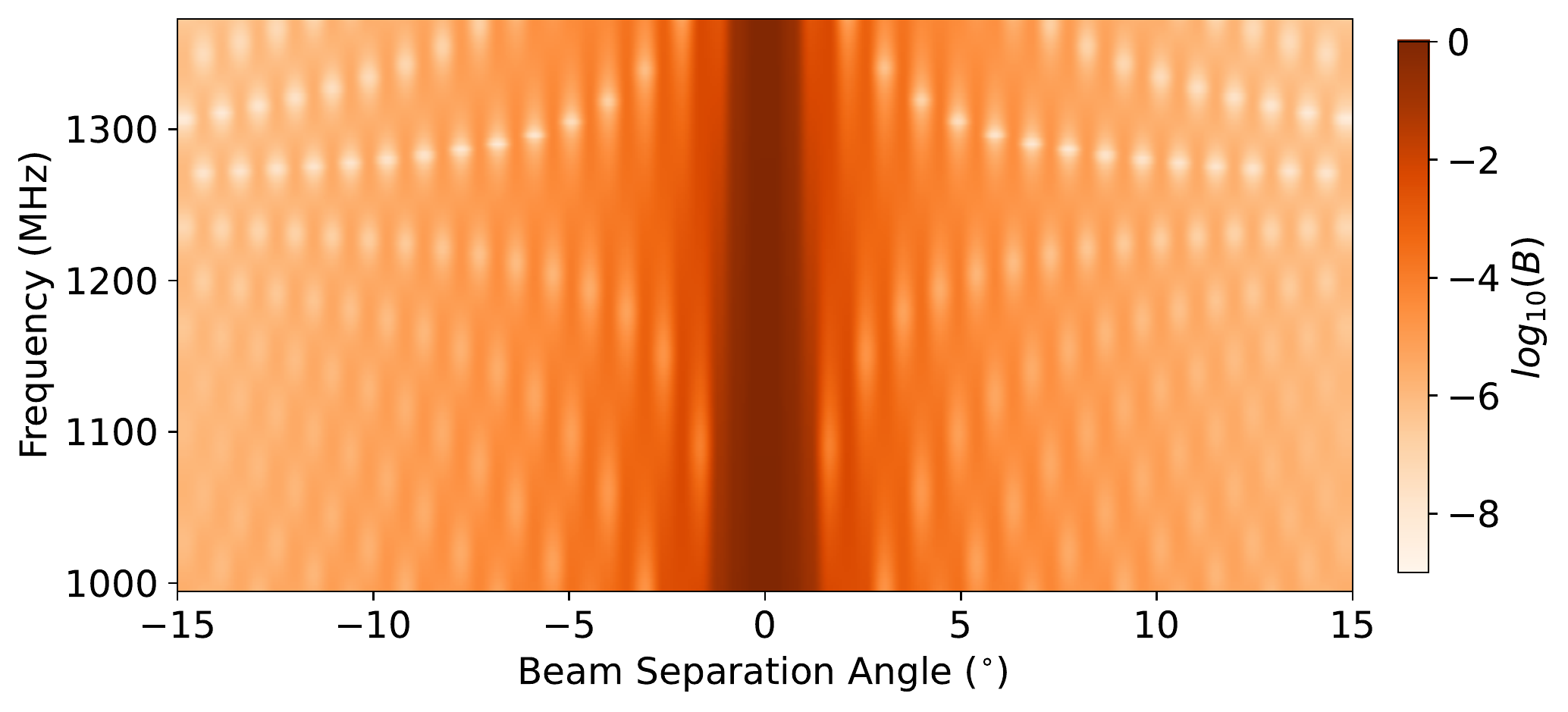}}\\
 \caption{A 1D slice through the \texttt{katbeam} beam model as a function of frequency.}
 \label{fig:beam}
   \end{figure}

By introducing high resolution spatial structure we aim to determine whether said structure poses a problem for component separation methods after the sky signal has been convolved with a complex beam structure, i.e. a frequency changing beam with sidelobes which cannot be modelled at each frequency as Gaussian. Through the \texttt{FastBox} set-up we make use of the L-Band, Stokes I \texttt{katbeam} model \footnote{\url{https://github.com/ska-sa/katbeam}} which models the frequency changing beam as a cosine aperture taper \citep{katbeam}. A 1D slice through the 2D beam pattern as a function of frequency is show in \autoref{fig:beam}. The approximate Gaussian FWHM for the MeerKAT beam for the frequency ranges under investigation in this work is 1.54$^{\circ}$ to 1.13$^{\circ}$.   

\begin{figure}
 \centering
  {\includegraphics[width=0.69\linewidth]{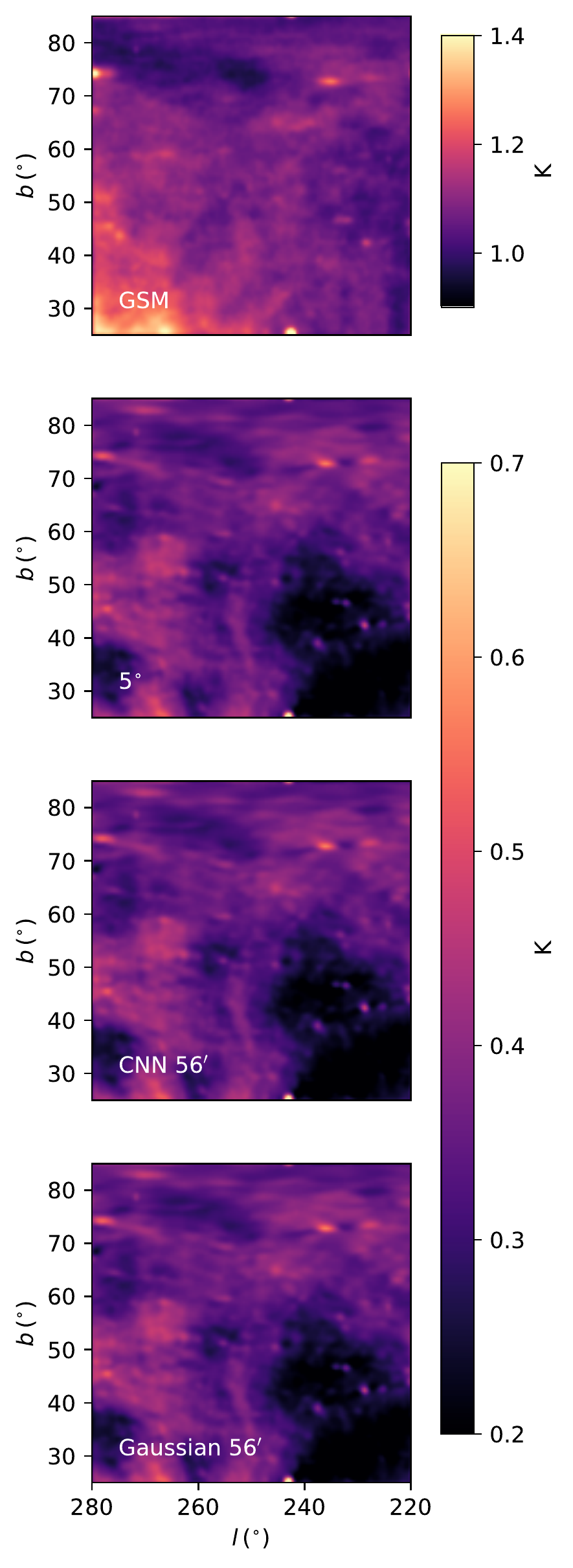}}\\
 \caption{The total emission models at 1273\,MHz for {\it{top to bottom:}} the GSM model, the $5^{\circ}$ model, the CNN model and the Gaussian model.}
 \label{fig:fgs}
   \end{figure}

Only the combination of the synchrotron emission plus the \hone \, signal is convolved with the beam; the final constituent of the total emission model: instrumental noise, is added after beam convolution. We assume Gaussian instrumental noise with a standard deviation calculated from the radiometer equation: 
\begin{equation}
\sigma_{{\rm{rms}}} = \frac{T_{{\rm{sys}}}} { \sqrt{N_{d} \, t_{{\rm{res}}} \, \delta \nu}}, 
\end{equation}
where
\begin{equation}
T_{{\rm{sys}}} = T_{r} + 60 \left( \frac{\nu}{300} \right)^{-2.5},
\end{equation}
where $T_{r} = 16$\,K (based on typical MeerKAT receiver temperatures \citep{wang} ), $N_d$ is the number of available dishes which we set to 64 to match the MeerKAT array, $\delta \nu$ is the frequency resolution and $t_{{\rm{res}}}$ is the observational time per pixel. \autoref{fig:fgs} shows the total emission (\hone \, plus synchrotron emission convolved with the beam and then added to the instrumental noise) at 1273\,MHz for the four different synchrotron emission models. The three models which use a scaling of the Haslam data for the foregrounds are indistinguishable by eye, whereas the total emission model which uses the GSM to provide the synchrotron emission template is quite distinct. All the models share common identifiable features however, such as regions of compact emission. 

\begin{figure}
 \centering
  {\includegraphics[width=0.99\linewidth]{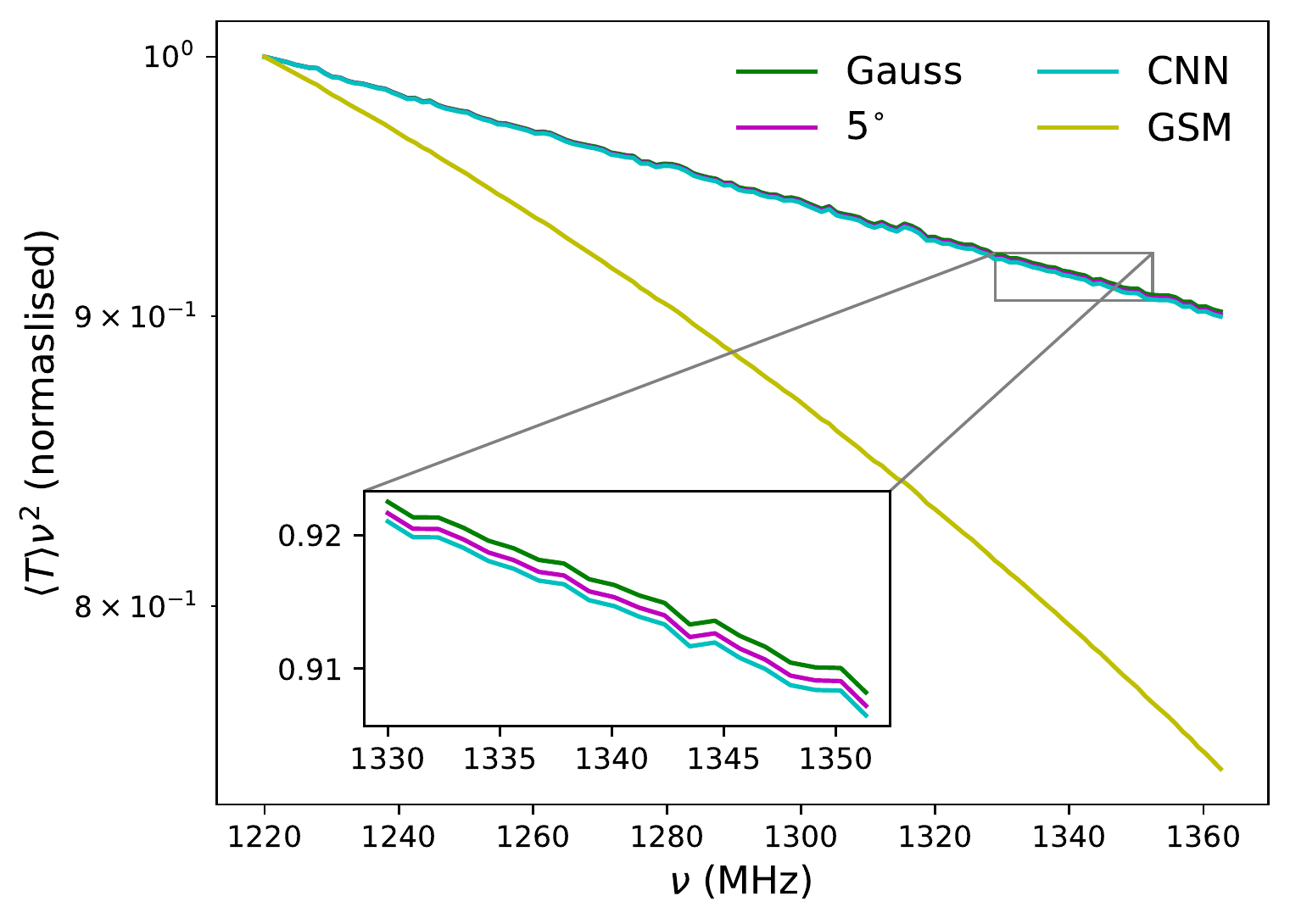}}\\
 \caption{The spectral form of a single pixel in each of the four total emission models.}
 \label{fig:wiggles}
   \end{figure}

In \autoref{fig:wiggles} the spectral form for a single pixel in each of the four total emission models is shown. The pixel temperature is multiplied by frequency squared in order to highlight any deviations from a simple power law model in the spectral form. The GSM total emission cube can be seen to display a completely different spectral form to the other three models, which is unsurprising as it is the only model not formed using a power law parameterisation. The other three models were formed from power laws but convolution with the MeerKAT beam has resulted in the averaging together of neighbouring pixels, which destroys the simple power law spectral form over frequency. The more complex the beam, the more complex these spectral perturbations and in-turn the harder it becomes to remove foreground structure when using a component separation technique that replies on spectral smoothness.    

 \subsection{The foreground clean}
Numerous methods of component separation can be used to attempt to clean total emission maps from foregrounds. Our aim is to determine if an individual technique behaves differently for our four sets of total emission cubes. Therefore we select a single method of foreground cleaning and explore the successes and failures of that technique for our four simulation sets. The method of choice for this work is Principal Component Analysis, which is a blind technique meaning that it requires no parametric information about the foreground emissions. PCA relies on the assumption that the foreground emission is smooth over frequency, whilst the cosmological signal is mainly stochastic over frequency. The technique works by performing an eigen decomposition of the data frequency-frequency covariance matrix; 
\begin{equation}
C(\nu, \nu^\prime) = \frac{1}{N_{\rm pix}} \sum_j^{N_{\rm pix}} \delta T_j(\nu) \delta T_j(\nu^\prime).
\end{equation}
The largest eigenmodes are then labelled as foregrounds and subtracted away from the total data. The number of these large eigenmodes to remove is the only input parameter required by the PCA algorithm from the user.     

\section{Results}
 \label{sec:res}
 
To asses the impact of our four different total emission models on the success of a PCA clean to the full data cube we enlist the use of spherically averaged auto-correlation power spectra. In place of plotting the power for each cube, we plot the ratio between the cleaned cube power spectrum and the \hone \, plus white noise power spectrum. An ideal foreground removal technique would remove all the synchrotron emission leaving only the \hone \, signal (after convolution with the beam) plus the instrumental noise thus giving a power spectra ratio of 1 on our plots. However, our focus is not on which foreground model results in the ratio closest to one but rather, if there are significant differences between the different total emission models.

\begin{figure}
 \centering
  {\includegraphics[width=0.99\linewidth]{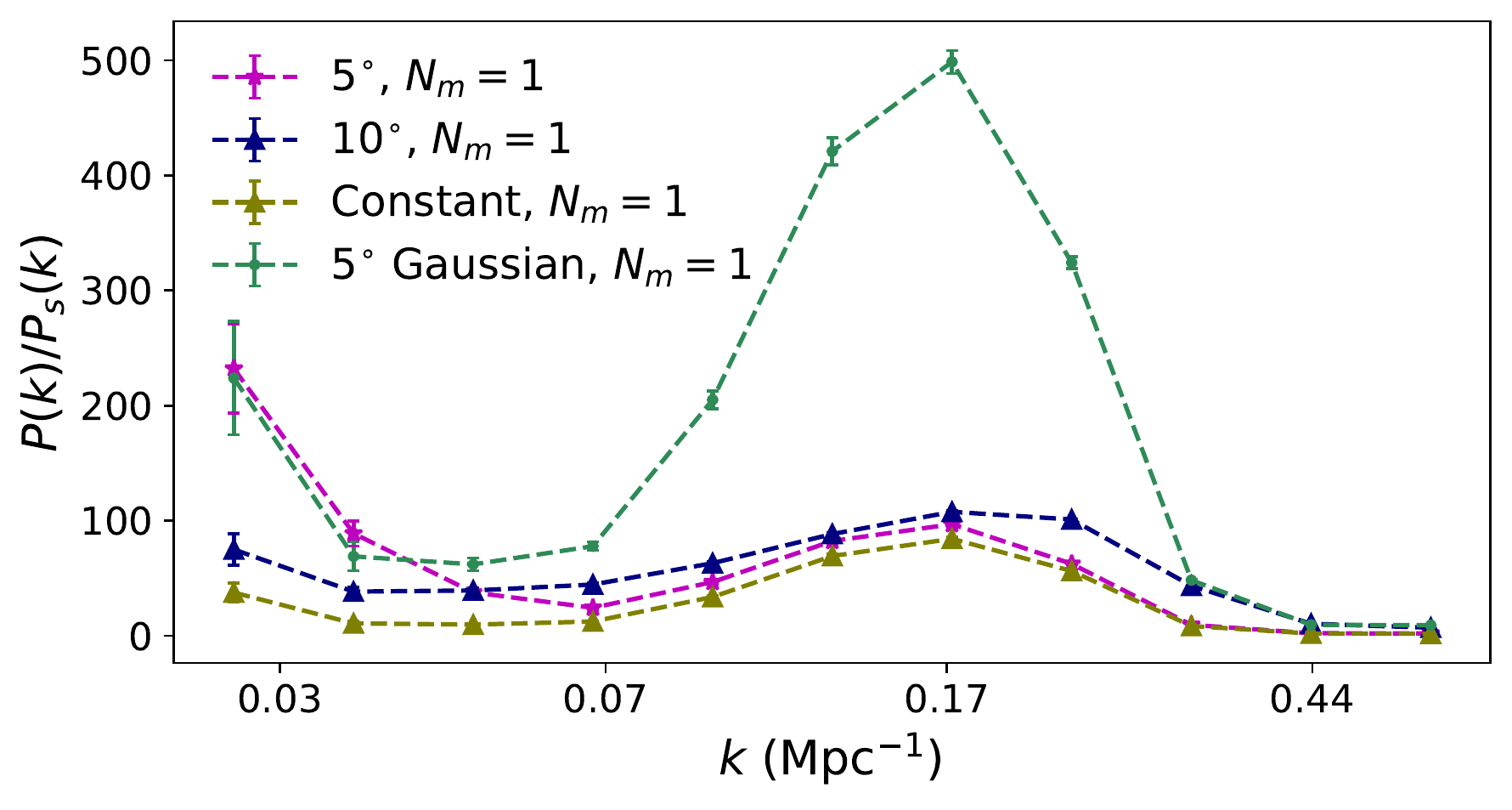}}\\
 \caption{Spherically averaged auto-correlation power spectra. The power spectra ratio between cleaned total emission cubes and data cubes containing only \hone \, emission convolved with the beam plus instrumental noise are plotted. In all four synchrotron emission models the Haslam data have been scaled using a spectral index map. The four different plots are for four different synchrotron spectral index maps.}
 \label{fig:thepoint}
   \end{figure}

In \autoref{fig:thepoint} we summarise the current understanding of synchrotron spectral index modelling for the problem of component separation. The auto-correlation power spectra shown are all for PCA-cleaned total emission cubes, the number of PCA modes removed from the total emission cubes is always 1. The synchrotron emission models for the four spectra were provided by the Haslam data scaled to different frequencies using four different spectral index maps. The first, and simplest spectral index map is simply a constant value at each pixel. We use the mean value of the 5 degree \citet{mamd} spectral index map, -2.93, as our constant spectral index value. The second map is a Gaussian distribution of values with a mean of -2.93 and the same standard deviation as the 5 degree spectral index map, 0.06, with a spatial resolution of 5 degrees. The third map is the actual 5 degree spectral index map and the fourth map is the third map smoothed to a resolution of 10 degrees. As previously identified in the literature, using a single, constant value for the synchrotron spectral index across all pixels presents a simplified version of the true problem to the component separation algorithm. The fact that the three 5 and 10 degree power spectra display the largest power excesses over the \hone \, plus white noise power indicates that the foreground removal is particularly challenging for a spatially complex synchrotron spectral index. The Gaussian 5 degree spectral index map, however, can be seen to place an excess of power across different (smaller) angular scales to the true 5 degree spectral index map and therefore is not an accurate representation of the foreground cleaning problem at medium angular scales. Whereas, the power spectra ratios with the largest values at large scales (small $k$ values), as opposed to medium scales, are the two where a non-Gaussian, spatially changing spectral index has been used. As there is a significant difference between the cleaned power spectra for the true 10 and 5 degree spectral index map this plot motivates the goal of this paper, which is to determine if a new spectral index map with non-Gaussian spatial structure and a resolution of 56 arcmin is required to further bolster \hone \, component separation test-beds.

 \begin{figure}
 \centering
  {\includegraphics[width=0.99\linewidth]{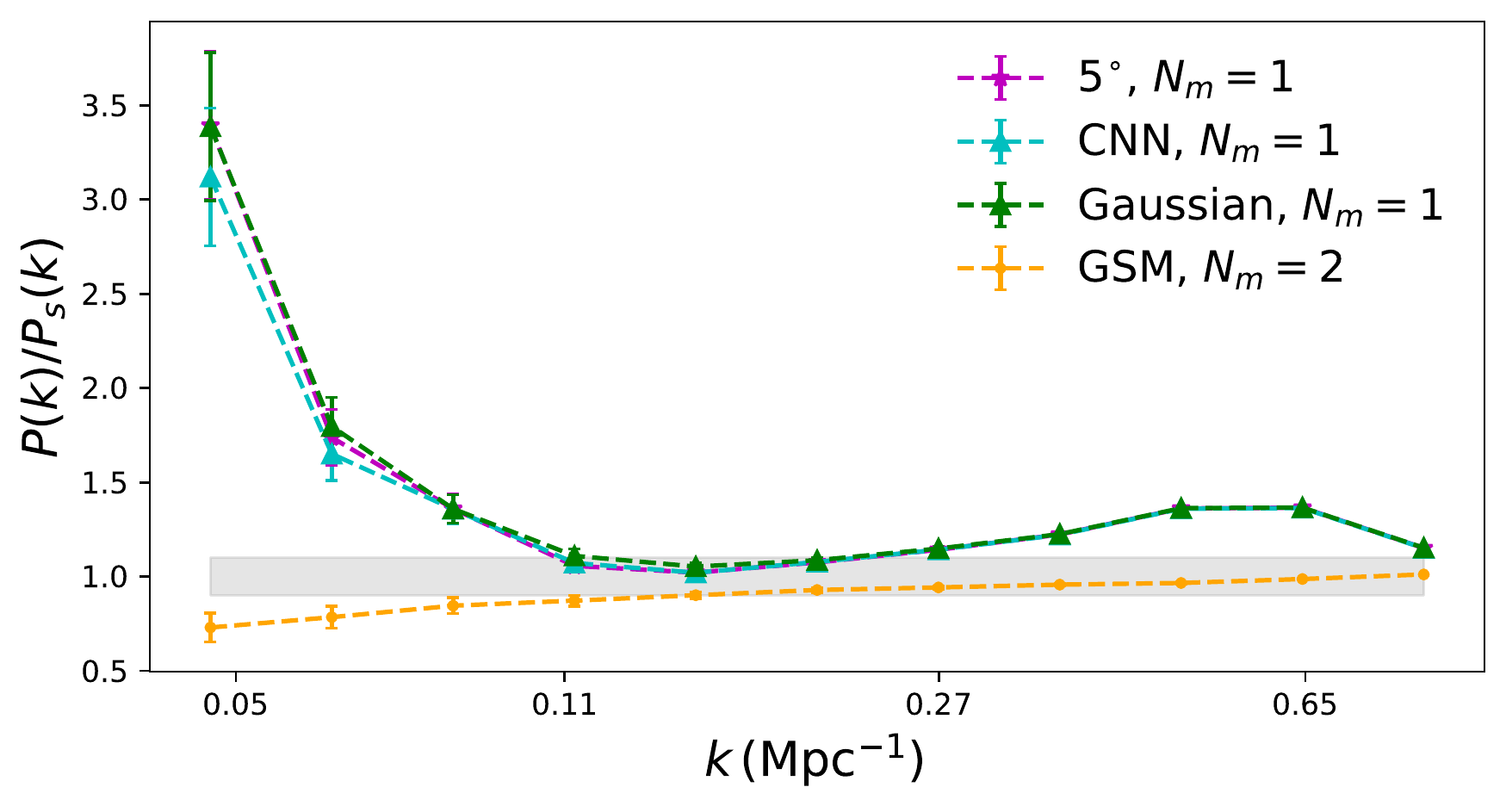}}\\
      {\includegraphics[width=0.99\linewidth]{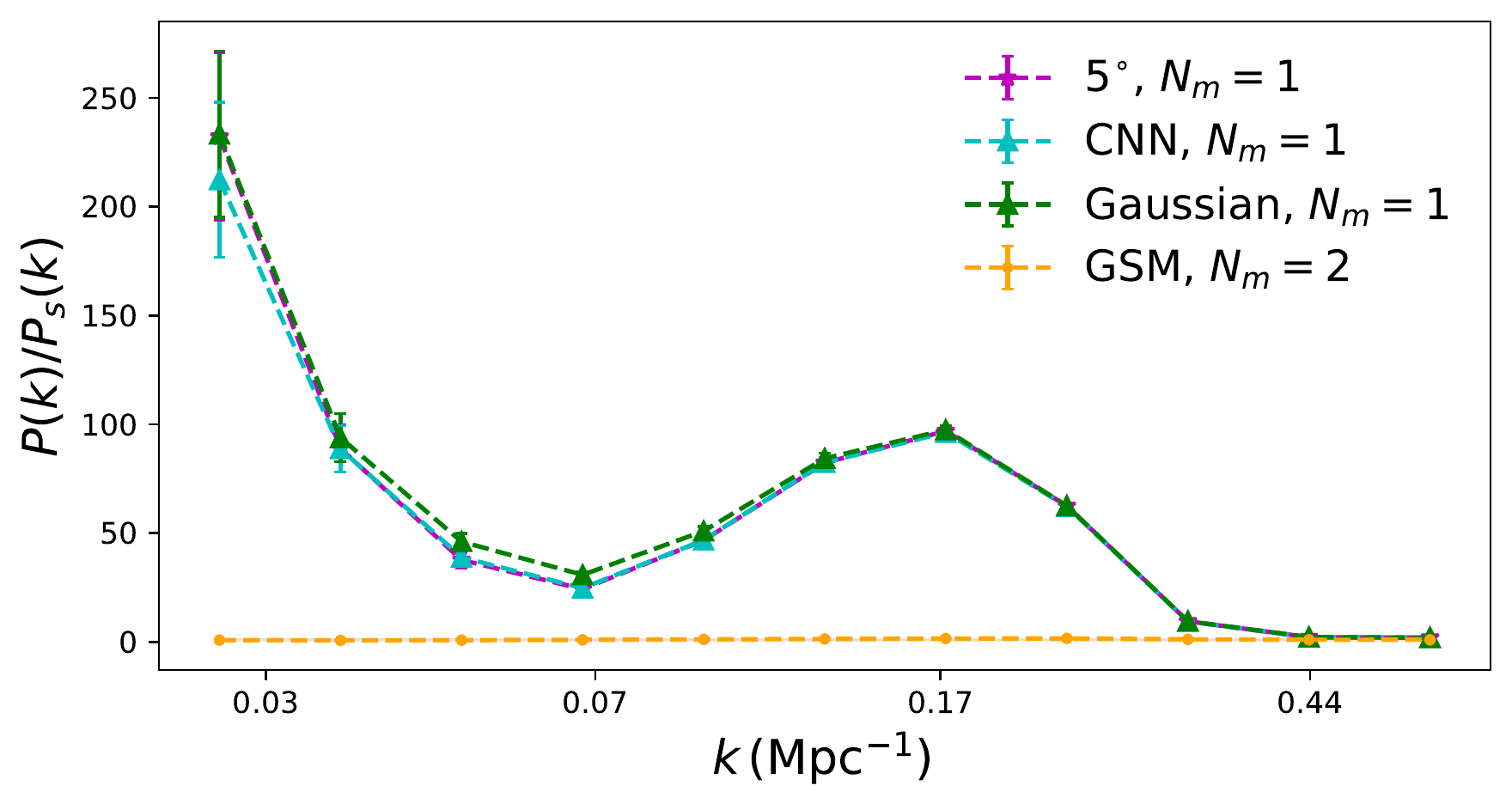}}\\
      {\includegraphics[width=0.99\linewidth]{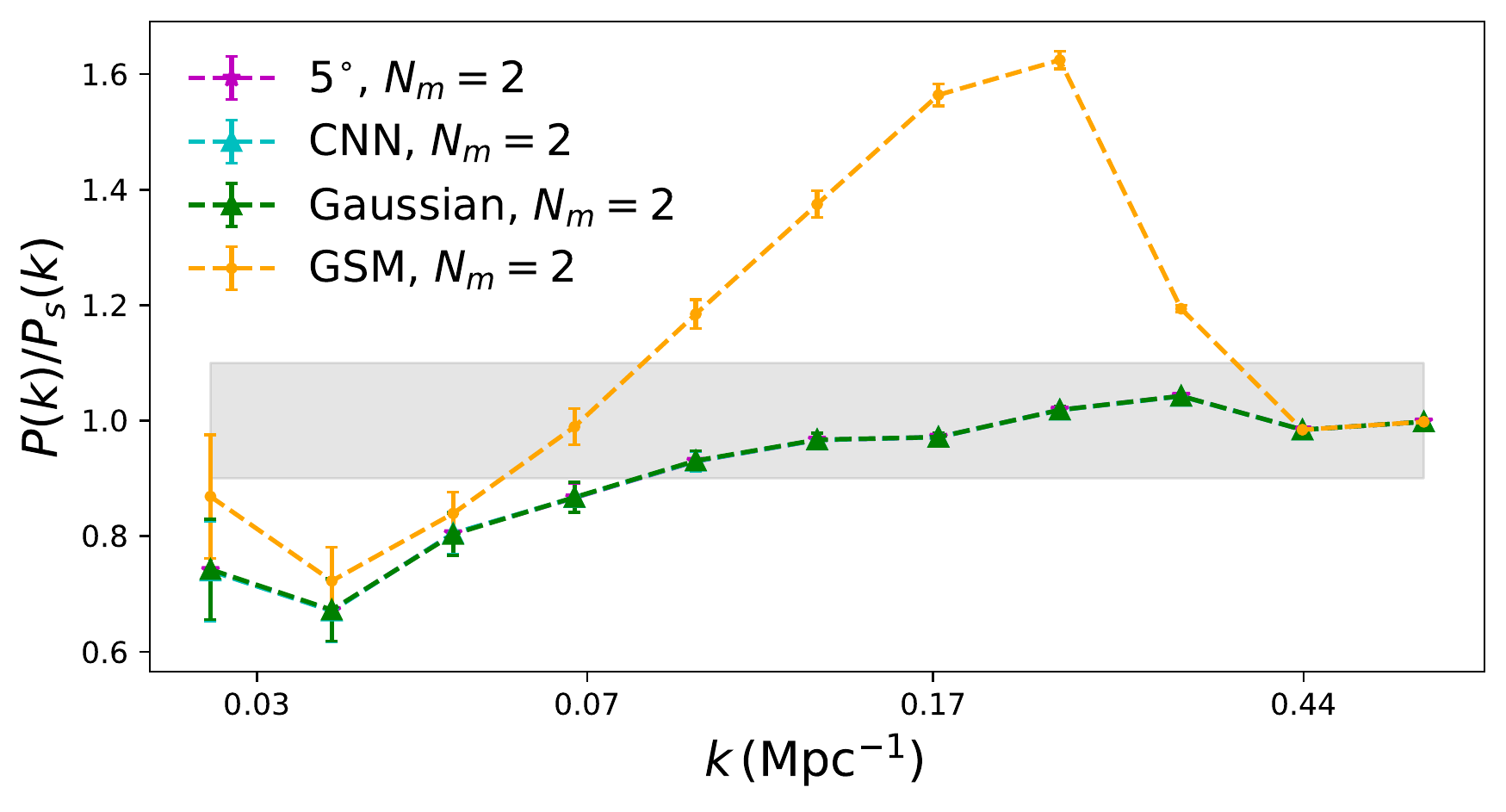}}\\
 \caption{Spherically averaged auto-correlation power spectra. The power spectra ratio between cleaned total emission cubes and data cubes containing only \hone \, emission convolved with the beam plus instrumental noise are plotted. The top plot is for total emission cubes covering the 1220.0 to 1362.6\,MHz frequency while the bottom two plots are for the 1084.0 to 1225.6\,MHz. The difference between the middle and the bottom plot is the number of foreground modes removed for the PCA clean.}
 \label{fig:1dps}
   \end{figure}
   
In the next plot we return to the four emission models set-up in \autoref{sec:theset}: namely the GSM, the 5 degree spectral index model, the 56 arcmin CNN spectral index model and the 56 arcmin Gaussian spectral index model. \autoref{fig:1dps} presents the auto-correlation power spectra plots for the four total emission models across the two frequency ranges investigated. The top plot is for the 1220.0 to 1362.6\,MHz frequency range whilst the bottom two are for the 1084.0 to 1225.6\,MHz frequency range. At lower frequencies the foreground emission is brighter and so a larger number of modes may be needed to achieve a suitably effective clean, which is why we include both the 1 and 2 mode removal results for the 1084.0 to 1225.6\,MHz range. The GSM total emission cube was only ever cleaned using 2 mode removal as 1 mode left visible foreground structure in the cleaned maps. It is also clear from the plots in \autoref{fig:1dps} that the GSM model results in a completely different cleaned result compared to the other three models which are formed from a power law extrapolation of the Haslam data. The GSM is a total diffuse emission model and therefore should be used as such, as opposed to as a proxy for purely synchrotron emission. It has a more complex spectral structure than a simple power law, which is an advantage to those who wish to test their component separation technique of choice on such a data set. However, for those who intend to add their own compact sources, free-free emission, residual RFI etc., using the GSM as a proxy for solely synchrotron emission may make for an unrealistically hard to clean total emission cube. 
   
   
The three models which are based on a power law extrapolation of the Haslam data can be seen to perform similarly in cleaning from \autoref{fig:1dps}. In fact for the 1084.0 to 1225.6\,MHz frequency range when two foreground modes are removed and the total emission cube is over-cleaned, i.e. \hone \, signal itself is removed causing the power spectra ratio to drop below one, the three power law models are essentially identical. For the 1 mode clean, for both frequency ranges investigated, we can see a difference in the three power law models at the large scale, low $k$ end of the power spectra. At the largest scale plotted the 5 degree model and the Gaussian model behave similarly, while the CNN model gives different results. For the first frequency range the cleaned maps are between 3 and 3.5 times higher than the ideal measured signal (\hone \, plus white noise), the difference between 3 and 3.5 is noteworthy, although within the error budget. However for the second range, all three methods are more than 100 times higher in power than the ideal signal and so the difference between the three methods becomes insignificant compared to the overall inability to recover the \hone \, signal. The conclusion to be taken from this is that it is worthwhile to pursue a more spatially complex foreground model, such as the CNN model presented in this work, if the component separation method under testing is performing well. If on the other hand, a new method is being honed which is orders of magnitude away from recovering the \hone \, signal, there is no need to pursue a more complex test-bed set-up than using the Haslam map extrapolated to different frequencies with the \citet{mamd} spectral index map. Specifically, for the MeerKLASS experimental set-up and the foreground cleaning method of PCA there is no significant difference between any of our 56 arcmin and the 5 degree spectral index models i.e. spectral index spatial complexity beyond a 5 degree resolution is not required for this particular test set-up.

\section{Conclusions}
 \label{sec:conc}
 
Prior to this work it was known that assuming a spatially constant synchrotron spectral index would be an unrealistic simplification of the component separation problem for 21\,cm intensity mapping experiments with frequency-dependant beams. It had also been shown that Gaussian foreground realisations are easier to separate from the \hone \, signal than non-Gaussian foregrounds when using blind foreground removal techniques. Motivated by the knowledge that increasingly spatially complex synchrotron spectral index models have been required to test the limitations of component separation techniques we created a 56 arcmin spectral index model to act as a proxy until future observational data allows us to measure the all-sky synchrotron spectral index at high angular resolution.

We have presented a model of the diffuse Galactic synchrotron emission spectral index between 0.408 and 23\,GHz created from the 5 degree map of \citet{mamd} using a CNN trained on low and high resolution spectral index maps from CHIPASS 1.4\,GHz observational data. Our map contains spatial structure up to 56 arcmin resolution. The intent is for this map to be used alongside the 56 arcmin 408\,MHz map to create more realistic models for diffuse synchrotron emission across frequency for use as part of any simulation suite designed to test component separation techniques. It can be seen from the test subset of the data used to train the CNN that the small-scale structure generated is not an accurate representation of the true (empirically measured) small-scale structure, this coupled with the mid-range resolution power loss seen in the angular power spectrum means that our spectral index map cannot be used for science analysis of diffuse Galactic synchrotron emission. However, we believe that the CNN model presented in this work offers a useful contribution to simulation test-beds designed to probe the advantages and disadvantages of different component separation techniques.    

Deep convolutional generative adversarial networks (DCGANs) have been implemented for the simulation of thermal dust emission maps {\citep{forse, aylor} but for our input resolution (5 degrees) and relatively small number of training maps (156) we found it problematic to get a GAN network to converge and so found instead, the U-NET CNN architecture to be the optimum set-up for our purposes. The availability of high resolution CHIPASS data enabled the calculation of spectral indices between 408 and 1400\,MHz at 56 arcmin which could then be smoothed so our network could be trained using pairs of 56 arcmin and 5 degree spectral index maps. Ideally we would have used publicly available high resolution MHz data, as at 1.4 GHz free-free emission is no longer negligible in certain regions of the sky. To combat the lack of available MHz data, we used the CHIPASS data but only within the North Polar Spur region; an area known to be dominated by diffuse synchrotron emission. In the future, however, we will be able to redo this analysis using MeerKAT or Bingo data to train our CNN.

To emphasize the use of a spatially complex foreground emission model, we set up four different total emission models and tested the ability of PCA to clean away the foregrounds, leaving the \hone \, plus instrumental noise. Our four emission models each contained the same \hone \, signal, the same level of instrumental Gaussian noise and the foreground and sky were convolved by the frequency-dependant \texttt{katbeam} beam model for each total emission model. The only difference between the four models were the foreground contributions. Emission model one had the synchrotron emission provided by the Global Sky Model, emission models two to four were all power law models of synchrotron emission made by extrapolating the Haslam map over frequency using a spectral index map. For model two the spectral index map used was the 5 degree map of \citet{mamd}, for model three our high resolution CNN spectral index map was used and for model four the high resolution spatial information for the 5 degree spectral index map was provided using a Gaussian realisation. None of these four models are perfect, the GSM model is for total emission and therefore has a far more complex spectral form than a simple power law and the other three models assume that only synchrotron emission is present in the Haslam data. Having a selection of possible models with different characteristics, however, is useful for testing component separation techniques as that allows for an investigation into how each technique responds to different contaminants. Using the GSM allows for the testing of a technique in the presence of a complex spectral structure caused by multiple emission sources, whilst using a power law model with a spatially varying spectral index simulates the complex spectral structure caused by the interaction between a foreground emission and the telescope beam pattern. For our experimental setup (cosine aperture taper beam with a FWHM between 1.1 and 1.6 degrees and foregrounds cleaned using PCA) and an upper resolution bound of 5 degrees} we have shown that increasing the resolution of the spectral index spatial structure changes the cleaned power spectrum and that this change occurs over different $k$ scales depending on whether the spatial structure is Gaussian or non-Gaussian. At resolutions equal to or greater than 5 degrees, however, there is no significant difference between using the \citet{mamd} 5 degree spectral index map, a Gaussian 56 arcmin spectral index map or a non-Gaussian 56 arcmin spectral index map. This case-study is of particular pertinence for the MeerKLASS \citep{MKwhite} component separation effort, while the 56 arcmin spectral index map made publicly available by this work is relevant to any other 21\,cm experiment testing component separation techniques on their own unique experimental set-up.

In this work we have chosen to use the \citet{mamd} all-sky map as our estimate for the synchrotron spectral index at all resolutions larger than 5 degrees. As of such our CNN spectral index map is completely tied to this per-pixel spectral index estimate. We believe this is an appropriate choice given the prevalence of this spectral index map within both the CMB and intensity mapping communities as it is used within both the Python Sky Model and Planck Sky Model. However, it must be noted that any limitations or inaccuracies associated with the \citet{mamd} 5 degree spectral index map are shared by the CNN 56 arcmin spectral index map presented in this paper.

An additional complexity for the synchrotron spectral index is the very likely possibility that it changes, not only across pixels but also over frequency. Our CNN spectral index map, like the 5 degree spectral index map, provides the average (across frequency) spectral index per pixel between 0.408 and 23\,GHz. The per-pixel spectral index value can, however, be scaled across frequency using a curvature model from the literature (\citet{kogut} for example), if required. We leave it to the community to scale the spectral index map as desired and to extend the tests shown in this work to include any other component separation technique and other non-Gaussian beam models. 
     
\section*{Data Availability}
The 56 arcmin map of the simulated synchrotron spectral index is publicly available here: \url{https://github.com/melisirfan/synchrotron_emission} and the \texttt{Jupyter} notebooks used in this analysis have been added to the \texttt{Fastbox} repository.

\section*{Acknowledgements}
M.I acknowledges support from the South African Radio Astronomy Observatory and National Research Foundation (Grant No. 84156) and would like to thank Mosima Masipa for her valuable insight on CNN architecture and Phil Bull and Mario Santos for the useful discussions.  

\balance


\bibliographystyle{mnras}
\bibliography{refs} 

\begin{thebibliography}{}
\makeatletter
\relax
\def\mn@urlcharsother{\let\do\@makeother \do\$\do\&\do\#\do\^\do\_\do\%\do\~}
\def\mn@doi{\begingroup\mn@urlcharsother \@ifnextchar [ {\mn@doi@}
  {\mn@doi@[]}}
\def\mn@doi@[#1]#2{\def\@tempa{#1}\ifx\@tempa\@empty \href
  {http://dx.doi.org/#2} {doi:#2}\else \href {http://dx.doi.org/#2} {#1}\fi
  \endgroup}
\def\mn@eprint#1#2{\mn@eprint@#1:#2::\@nil}
\def\mn@eprint@arXiv#1{\href {http://arxiv.org/abs/#1} {{\tt arXiv:#1}}}
\def\mn@eprint@dblp#1{\href {http://dblp.uni-trier.de/rec/bibtex/#1.xml}
  {dblp:#1}}
\def\mn@eprint@#1:#2:#3:#4\@nil{\def\@tempa {#1}\def\@tempb {#2}\def\@tempc
  {#3}\ifx \@tempc \@empty \let \@tempc \@tempb \let \@tempb \@tempa \fi \ifx
  \@tempb \@empty \def\@tempb {arXiv}\fi \@ifundefined
  {mn@eprint@\@tempb}{\@tempb:\@tempc}{\expandafter \expandafter \csname
  mn@eprint@\@tempb\endcsname \expandafter{\@tempc}}}

\bibitem[\protect\citeauthoryear{{Alonso}, {Bull}, {Ferreira}  \&
  {Santos}}{{Alonso} et~al.}{2015}]{alonsoFGsim}
{Alonso} D.,  {Bull} P.,  {Ferreira} P.~G.,   {Santos} M.~G.,  2015, \mn@doi
  [\mnras] {10.1093/mnras/stu2474}, \href
  {https://ui.adsabs.harvard.edu/abs/2015MNRAS.447..400A} {447, 400}

\bibitem[\protect\citeauthoryear{{Anstey}, {de Lera Acedo}  \&
  {Handley}}{{Anstey} et~al.}{2021}]{anstey}
{Anstey} D.,  {de Lera Acedo} E.,   {Handley} W.,  2021, \mn@doi [\mnras]
  {10.1093/mnras/stab1765}, \href
  {https://ui.adsabs.harvard.edu/abs/2021MNRAS.506.2041A} {506, 2041}

\bibitem[\protect\citeauthoryear{{Aylor}, {Haq}, {Knox}, {Hezaveh}  \&
  {Perreault-Levasseur}}{{Aylor} et~al.}{2021}]{aylor}
{Aylor} K.,  {Haq} M.,  {Knox} L.,  {Hezaveh} Y.,   {Perreault-Levasseur} L.,
  2021, \mn@doi [\mnras] {10.1093/mnras/staa3344}, \href
  {https://ui.adsabs.harvard.edu/abs/2021MNRAS.500.3889A} {500, 3889}

\bibitem[\protect\citeauthoryear{{Battye} et~al.,}{{Battye}
  et~al.}{2012}]{bingo}
{Battye} R.~A.,  et~al., 2012, arXiv e-prints, \href
  {https://ui.adsabs.harvard.edu/abs/2012arXiv1209.1041B} {p. arXiv:1209.1041}

\bibitem[\protect\citeauthoryear{{Bennett} et~al.,}{{Bennett}
  et~al.}{1992}]{ffbeta}
{Bennett} C.~L.,  et~al., 1992, \mn@doi [\apjl] {10.1086/186505}, \href
  {https://ui.adsabs.harvard.edu/abs/1992ApJ...396L...7B} {396, L7}

\bibitem[\protect\citeauthoryear{{Bennett} et~al.,}{{Bennett}
  et~al.}{2003}]{ben03}
{Bennett} C.~L.,  et~al., 2003, \mn@doi [\apjs] {10.1086/377252}, \href
  {https://ui.adsabs.harvard.edu/abs/2003ApJS..148...97B} {148, 97}

\bibitem[\protect\citeauthoryear{{Berkhuijsen}}{{Berkhuijsen}}{1972}]{dwing}
{Berkhuijsen} E.~M.,  1972, \aaps, \href
  {https://ui.adsabs.harvard.edu/abs/1972A&AS....5..263B} {5, 263}

\bibitem[\protect\citeauthoryear{{Bernardi}, {McQuinn}  \&
  {Greenhill}}{{Bernardi} et~al.}{2015}]{bern15}
{Bernardi} G.,  {McQuinn} M.,   {Greenhill} L.~J.,  2015, \mn@doi [\apj]
  {10.1088/0004-637X/799/1/90}, \href
  {https://ui.adsabs.harvard.edu/abs/2015ApJ...799...90B} {799, 90}

\bibitem[\protect\citeauthoryear{{Bigot-Sazy} et~al.,}{{Bigot-Sazy}
  et~al.}{2015}]{bigotFGsim}
{Bigot-Sazy} M.~A.,  et~al., 2015, \mn@doi [\mnras] {10.1093/mnras/stv2153},
  \href {https://ui.adsabs.harvard.edu/abs/2015MNRAS.454.3240B} {454, 3240}

\bibitem[\protect\citeauthoryear{{Calabretta}, {Staveley-Smith}  \&
  {Barnes}}{{Calabretta} et~al.}{2014}]{chipass}
{Calabretta} M.~R.,  {Staveley-Smith} L.,   {Barnes} D.~G.,  2014, \mn@doi
  [\pasa] {10.1017/pasa.2013.36}, \href
  {https://ui.adsabs.harvard.edu/abs/2014PASA...31....7C} {31, e007}

\bibitem[\protect\citeauthoryear{{Carucci}, {Irfan}  \& {Bobin}}{{Carucci}
  et~al.}{2020}]{isaFGsim}
{Carucci} I.~P.,  {Irfan} M.~O.,   {Bobin} J.,  2020, \mn@doi [\mnras]
  {10.1093/mnras/staa2854}, \href
  {https://ui.adsabs.harvard.edu/abs/2020MNRAS.499..304C} {499, 304}

\bibitem[\protect\citeauthoryear{{Caswell}}{{Caswell}}{1976}]{drao}
{Caswell} J.~L.,  1976, \mn@doi [\mnras] {10.1093/mnras/177.3.601}, \href
  {https://ui.adsabs.harvard.edu/abs/1976MNRAS.177..601C} {177, 601}

\bibitem[\protect\citeauthoryear{{Chapman}, {Zaroubi}, {Abdalla}, {Dulwich},
  {Jeli{\'c}}  \& {Mort}}{{Chapman} et~al.}{2016}]{chapFGsim}
{Chapman} E.,  {Zaroubi} S.,  {Abdalla} F.~B.,  {Dulwich} F.,  {Jeli{\'c}} V.,
   {Mort} B.,  2016, \mn@doi [\mnras] {10.1093/mnras/stw161}, \href
  {https://ui.adsabs.harvard.edu/abs/2016MNRAS.458.2928C} {458, 2928}

\bibitem[\protect\citeauthoryear{Coifman \& Donoho}{Coifman \&
  Donoho}{1995}]{cycle}
Coifman R.~R.,  Donoho D.~L.,  1995, Translation-Invariant De-Noising.
Springer New York, New York, NY, pp 125--150,
  \mn@doi{10.1007/978-1-4612-2544-7_9}, \url
  {https://doi.org/10.1007/978-1-4612-2544-7_9}

\bibitem[\protect\citeauthoryear{{Cunnington}, {Irfan}, {Carucci}, {Pourtsidou}
   \& {Bobin}}{{Cunnington} et~al.}{2021}]{steveFGsim}
{Cunnington} S.,  {Irfan} M.~O.,  {Carucci} I.~P.,  {Pourtsidou} A.,   {Bobin}
  J.,  2021, \mn@doi [\mnras] {10.1093/mnras/stab856}, \href
  {https://ui.adsabs.harvard.edu/abs/2021MNRAS.504..208C} {504, 208}

\bibitem[\protect\citeauthoryear{{DeBoer} et~al.,}{{DeBoer}
  et~al.}{2017}]{hera}
{DeBoer} D.~R.,  et~al., 2017, \mn@doi [\pasp]
  {10.1088/1538-3873/129/974/045001}, \href
  {https://ui.adsabs.harvard.edu/abs/2017PASP..129d5001D} {129, 045001}

\bibitem[\protect\citeauthoryear{{Delabrouille} et~al.,}{{Delabrouille}
  et~al.}{2013}]{psm}
{Delabrouille} J.,  et~al., 2013, \mn@doi [\aap] {10.1051/0004-6361/201220019},
  \href {https://ui.adsabs.harvard.edu/abs/2013A&A...553A..96D} {553, A96}

\bibitem[\protect\citeauthoryear{{Dickinson} et~al.,}{{Dickinson}
  et~al.}{2009}]{clive09}
{Dickinson} C.,  et~al., 2009, \mn@doi [\apj] {10.1088/0004-637X/705/2/1607},
  \href {https://ui.adsabs.harvard.edu/abs/2009ApJ...705.1607D} {705, 1607}

\bibitem[\protect\citeauthoryear{{Eastwood} et~al.,}{{Eastwood}
  et~al.}{2019}]{lwa}
{Eastwood} M.~W.,  et~al., 2019, \mn@doi [\aj] {10.3847/1538-3881/ab2629},
  \href {https://ui.adsabs.harvard.edu/abs/2019AJ....158...84E} {158, 84}

\bibitem[\protect\citeauthoryear{{Fauvet} et~al.,}{{Fauvet}
  et~al.}{2011}]{magfield1}
{Fauvet} L.,  et~al., 2011, \mn@doi [\aap] {10.1051/0004-6361/201014492}, \href
  {https://ui.adsabs.harvard.edu/abs/2011A&A...526A.145F} {526, A145}

\bibitem[\protect\citeauthoryear{{Gold} et~al.,}{{Gold} et~al.}{2009}]{gold09}
{Gold} B.,  et~al., 2009, \mn@doi [\apjs] {10.1088/0067-0049/180/2/265}, \href
  {https://ui.adsabs.harvard.edu/abs/2009ApJS..180..265G} {180, 265}

\bibitem[\protect\citeauthoryear{{Gold} et~al.,}{{Gold} et~al.}{2011}]{gold11}
{Gold} B.,  et~al., 2011, \mn@doi [\apjs] {10.1088/0067-0049/192/2/15}, \href
  {https://ui.adsabs.harvard.edu/abs/2011ApJS..192...15G} {192, 15}

\bibitem[\protect\citeauthoryear{{G{\'o}rski}, {Hivon}, {Banday}, {Wandelt},
  {Hansen}, {Reinecke}  \& {Bartelmann}}{{G{\'o}rski} et~al.}{2005}]{healpix}
{G{\'o}rski} K.~M.,  {Hivon} E.,  {Banday} A.~J.,  {Wandelt} B.~D.,  {Hansen}
  F.~K.,  {Reinecke} M.,   {Bartelmann} M.,  2005, \mn@doi [\apj]
  {10.1086/427976}, \href
  {https://ui.adsabs.harvard.edu/abs/2005ApJ...622..759G} {622, 759}

\bibitem[\protect\citeauthoryear{{Guzm{\'a}n}, {May}, {Alvarez}  \&
  {Maeda}}{{Guzm{\'a}n} et~al.}{2011}]{mu}
{Guzm{\'a}n} A.~E.,  {May} J.,  {Alvarez} H.,   {Maeda} K.,  2011, \mn@doi
  [\aap] {10.1051/0004-6361/200913628}, \href
  {https://ui.adsabs.harvard.edu/abs/2011A&A...525A.138G} {525, A138}

\bibitem[\protect\citeauthoryear{{Haslam}, {Large}  \& {Quigley}}{{Haslam}
  et~al.}{1964}]{nps}
{Haslam} C.~G.~T.,  {Large} M.~I.,   {Quigley} M.~J.~S.,  1964, \mn@doi
  [\mnras] {10.1093/mnras/127.4.273}, \href
  {https://ui.adsabs.harvard.edu/abs/1964MNRAS.127..273H} {127, 273}

\bibitem[\protect\citeauthoryear{{Haslam}, {Salter}, {Stoffel}  \&
  {Wilson}}{{Haslam} et~al.}{1982}]{haslam}
{Haslam} C.~G.~T.,  {Salter} C.~J.,  {Stoffel} H.,   {Wilson} W.~E.,  1982,
  \aaps, \href {https://ui.adsabs.harvard.edu/abs/1982A&AS...47....1H} {47, 1}

\bibitem[\protect\citeauthoryear{{Irfan} \& {Bull}}{{Irfan} \&
  {Bull}}{2021}]{kpca}
{Irfan} M.~O.,  {Bull} P.,  2021, \mn@doi [\mnras] {10.1093/mnras/stab2855},
  \href {https://ui.adsabs.harvard.edu/abs/2021MNRAS.508.3551I} {508, 3551}

\bibitem[\protect\citeauthoryear{{Irfan} et~al.,}{{Irfan} et~al.}{2022}]{beta}
{Irfan} M.~O.,  et~al., 2022, \mn@doi [\mnras] {10.1093/mnras/stab3346}, \href
  {https://ui.adsabs.harvard.edu/abs/2022MNRAS.509.4923I} {509, 4923}

\bibitem[\protect\citeauthoryear{{Kogut}}{{Kogut}}{2012}]{kogut}
{Kogut} A.,  2012, \mn@doi [\apj] {10.1088/0004-637X/753/2/110}, \href
  {https://ui.adsabs.harvard.edu/abs/2012ApJ...753..110K} {753, 110}

\bibitem[\protect\citeauthoryear{{Krachmalnicoff} \&
  {Puglisi}}{{Krachmalnicoff} \& {Puglisi}}{2021}]{forse}
{Krachmalnicoff} N.,  {Puglisi} G.,  2021, \mn@doi [\apj]
  {10.3847/1538-4357/abe71c}, \href
  {https://ui.adsabs.harvard.edu/abs/2021ApJ...911...42K} {911, 42}

\bibitem[\protect\citeauthoryear{Kriele, Wayth, Bentum, Juswardy  \&
  Trott}{Kriele et~al.}{2022}]{eda}
Kriele M.~A.,  Wayth R.~B.,  Bentum M.~J.,  Juswardy B.,   Trott C.~M.,  2022,
  \mn@doi [Publications of the Astronomical Society of Australia]
  {10.1017/pasa.2022.2}, 39, e017

\bibitem[\protect\citeauthoryear{{Landecker} \& {Wielebinski}}{{Landecker} \&
  {Wielebinski}}{1970}]{parkes}
{Landecker} T.~L.,  {Wielebinski} R.,  1970, Australian Journal of Physics
  Astrophysical Supplement, \href
  {https://ui.adsabs.harvard.edu/abs/1970AuJPA..16....1L} {16, 1}

\bibitem[\protect\citeauthoryear{{Leach} et~al.,}{{Leach}
  et~al.}{2008}]{leach08}
{Leach} S.~M.,  et~al., 2008, \mn@doi [\aap] {10.1051/0004-6361:200810116},
  \href {https://ui.adsabs.harvard.edu/abs/2008A&A...491..597L} {491, 597}

\bibitem[\protect\citeauthoryear{{Liccardo} et~al.,}{{Liccardo}
  et~al.}{2021}]{bingoFGsim}
{Liccardo} V.,  et~al., 2021, arXiv e-prints, \href
  {https://ui.adsabs.harvard.edu/abs/2021arXiv210701636L} {p. arXiv:2107.01636}

\bibitem[\protect\citeauthoryear{{Makinen}, {Lancaster}, {Villaescusa-Navarro},
  {Melchior}, {Ho}, {Perreault-Levasseur}  \& {Spergel}}{{Makinen}
  et~al.}{2021}]{deepFGsim}
{Makinen} T.~L.,  {Lancaster} L.,  {Villaescusa-Navarro} F.,  {Melchior} P.,
  {Ho} S.,  {Perreault-Levasseur} L.,   {Spergel} D.~N.,  2021, \mn@doi [\jcap]
  {10.1088/1475-7516/2021/04/081}, \href
  {https://ui.adsabs.harvard.edu/abs/2021JCAP...04..081M} {2021, 081}

\bibitem[\protect\citeauthoryear{{Mauch} et~al.,}{{Mauch}
  et~al.}{2020}]{katbeam}
{Mauch} T.,  et~al., 2020, \mn@doi [\apj] {10.3847/1538-4357/ab5d2d}, \href
  {https://ui.adsabs.harvard.edu/abs/2020ApJ...888...61M} {888, 61}

\bibitem[\protect\citeauthoryear{{Mertens}, {Ghosh}  \& {Koopmans}}{{Mertens}
  et~al.}{2018}]{mertFGsim}
{Mertens} F.~G.,  {Ghosh} A.,   {Koopmans} L.~V.~E.,  2018, \mn@doi [\mnras]
  {10.1093/mnras/sty1207}, \href
  {https://ui.adsabs.harvard.edu/abs/2018MNRAS.478.3640M} {478, 3640}

\bibitem[\protect\citeauthoryear{{Miville-Desch{\^e}nes}, {Ysard}, {Lavabre},
  {Ponthieu}, {Mac{\'\i}as-P{\'e}rez}, {Aumont}  \&
  {Bernard}}{{Miville-Desch{\^e}nes} et~al.}{2008}]{mamd}
{Miville-Desch{\^e}nes} M.~A.,  {Ysard} N.,  {Lavabre} A.,  {Ponthieu} N.,
  {Mac{\'\i}as-P{\'e}rez} J.~F.,  {Aumont} J.,   {Bernard} J.~P.,  2008,
  \mn@doi [\aap] {10.1051/0004-6361:200809484}, \href
  {https://ui.adsabs.harvard.edu/abs/2008A&A...490.1093M} {490, 1093}

\bibitem[\protect\citeauthoryear{{Mozdzen}, {Bowman}, {Monsalve}  \&
  {Rogers}}{{Mozdzen} et~al.}{2016}]{moz16}
{Mozdzen} T.~J.,  {Bowman} J.~D.,  {Monsalve} R.~A.,   {Rogers} A.~E.~E.,
  2016, \mn@doi [\mnras] {10.1093/mnras/stv2601}, \href
  {https://ui.adsabs.harvard.edu/abs/2016MNRAS.455.3890M} {455, 3890}

\bibitem[\protect\citeauthoryear{{Nan} et~al.,}{{Nan} et~al.}{2011}]{fast}
{Nan} R.,  et~al., 2011, \mn@doi [International Journal of Modern Physics D]
  {10.1142/S0218271811019335}, \href
  {https://ui.adsabs.harvard.edu/abs/2011IJMPD..20..989N} {20, 989}

\bibitem[\protect\citeauthoryear{{Newburgh} et~al.,}{{Newburgh}
  et~al.}{2014}]{chime}
{Newburgh} L.~B.,  et~al., 2014, {Calibrating CHIME: a new radio interferometer
  to probe dark energy}.
p. 91454V, \mn@doi{10.1117/12.2056962}

\bibitem[\protect\citeauthoryear{{Parsons} et~al.,}{{Parsons}
  et~al.}{2010}]{paper}
{Parsons} A.~R.,  et~al., 2010, \mn@doi [\aj] {10.1088/0004-6256/139/4/1468},
  \href {https://ui.adsabs.harvard.edu/abs/2010AJ....139.1468P} {139, 1468}

\bibitem[\protect\citeauthoryear{{Planck Collaboration} et~al.,}{{Planck
  Collaboration} et~al.}{2014}]{planck2014}
{Planck Collaboration} et~al., 2014, \mn@doi [\aap]
  {10.1051/0004-6361/201321580}, \href
  {https://ui.adsabs.harvard.edu/abs/2014A&A...571A..12P} {571, A12}

\bibitem[\protect\citeauthoryear{{Planck Collaboration} et~al.,}{{Planck
  Collaboration} et~al.}{2016a}]{planck16}
{Planck Collaboration} et~al., 2016a, \mn@doi [\aap]
  {10.1051/0004-6361/201525967}, \href
  {https://ui.adsabs.harvard.edu/abs/2016A&A...594A..10P} {594, A10}

\bibitem[\protect\citeauthoryear{{Planck Collaboration} et~al.,}{{Planck
  Collaboration} et~al.}{2016b}]{ffp}
{Planck Collaboration} et~al., 2016b, \mn@doi [\aap]
  {10.1051/0004-6361/201527103}, \href
  {https://ui.adsabs.harvard.edu/abs/2016A&A...594A..12P} {594, A12}

\bibitem[\protect\citeauthoryear{{Planck Collaboration} et~al.,}{{Planck
  Collaboration} et~al.}{2016c}]{gnilcDust}
{Planck Collaboration} et~al., 2016c, \mn@doi [\aap]
  {10.1051/0004-6361/201629022}, \href
  {https://ui.adsabs.harvard.edu/abs/2016A&A...596A.109P} {596, A109}

\bibitem[\protect\citeauthoryear{{Planck Collaboration} et~al.,}{{Planck
  Collaboration} et~al.}{2020}]{planck18}
{Planck Collaboration} et~al., 2020, \mn@doi [\aap]
  {10.1051/0004-6361/201833881}, \href
  {https://ui.adsabs.harvard.edu/abs/2020A&A...641A...4P} {641, A4}

\bibitem[\protect\citeauthoryear{{Platania}, {Bensadoun}, {Bersanelli}, {De
  Amici}, {Kogut}, {Levin}, {Maino}  \& {Smoot}}{{Platania}
  et~al.}{1998}]{plat}
{Platania} P.,  {Bensadoun} M.,  {Bersanelli} M.,  {De Amici} G.,  {Kogut} A.,
  {Levin} S.,  {Maino} D.,   {Smoot} G.~F.,  1998, \mn@doi [\apj]
  {10.1086/306175}, \href
  {https://ui.adsabs.harvard.edu/abs/1998ApJ...505..473P} {505, 473}

\bibitem[\protect\citeauthoryear{{Reich} \& {Reich}}{{Reich} \&
  {Reich}}{1986}]{stock}
{Reich} P.,  {Reich} W.,  1986, \aaps, \href
  {https://ui.adsabs.harvard.edu/abs/1986A&AS...63..205R} {63, 205}

\bibitem[\protect\citeauthoryear{{Remazeilles}, {Dickinson}, {Banday},
  {Bigot-Sazy}  \& {Ghosh}}{{Remazeilles} et~al.}{2015}]{destriped}
{Remazeilles} M.,  {Dickinson} C.,  {Banday} A.~J.,  {Bigot-Sazy} M.~A.,
  {Ghosh} T.,  2015, \mn@doi [\mnras] {10.1093/mnras/stv1274}, \href
  {https://ui.adsabs.harvard.edu/abs/2015MNRAS.451.4311R} {451, 4311}

\bibitem[\protect\citeauthoryear{{Roger}, {Costain}, {Landecker}  \&
  {Swerdlyk}}{{Roger} et~al.}{1999}]{drao2}
{Roger} R.~S.,  {Costain} C.~H.,  {Landecker} T.~L.,   {Swerdlyk} C.~M.,  1999,
  \mn@doi [\aaps] {10.1051/aas:1999239}, \href
  {https://ui.adsabs.harvard.edu/abs/1999A&AS..137....7R} {137, 7}

\bibitem[\protect\citeauthoryear{{Ronneberger}, {Fischer}  \&
  {Brox}}{{Ronneberger} et~al.}{2015}]{unet}
{Ronneberger} O.,  {Fischer} P.,   {Brox} T.,  2015, arXiv e-prints, \href
  {https://ui.adsabs.harvard.edu/abs/2015arXiv150504597R} {p. arXiv:1505.04597}

\bibitem[\protect\citeauthoryear{{Santos} \& {Cooray}}{{Santos} \&
  {Cooray}}{2006}]{santos05}
{Santos} M.~G.,  {Cooray} A.,  2006, \mn@doi [\prd]
  {10.1103/PhysRevD.74.083517}, \href
  {https://ui.adsabs.harvard.edu/abs/2006PhRvD..74h3517S} {74, 083517}

\bibitem[\protect\citeauthoryear{{Santos} et~al.,}{{Santos}
  et~al.}{2016}]{MKwhite}
{Santos} M.,  et~al., 2016, in MeerKAT Science: On the Pathway to the SKA.
  p.~32 (\mn@eprint {arXiv} {1709.06099})

\bibitem[\protect\citeauthoryear{{Shaw}, {Sigurdson}, {Sitwell}, {Stebbins}  \&
  {Pen}}{{Shaw} et~al.}{2015}]{shawFGsim}
{Shaw} J.~R.,  {Sigurdson} K.,  {Sitwell} M.,  {Stebbins} A.,   {Pen} U.-L.,
  2015, \mn@doi [\prd] {10.1103/PhysRevD.91.083514}, \href
  {https://ui.adsabs.harvard.edu/abs/2015PhRvD..91h3514S} {91, 083514}

\bibitem[\protect\citeauthoryear{{Soares}, {Watkinson}, {Cunnington}  \&
  {Pourtsidou}}{{Soares} et~al.}{2022}]{paulaFGsim}
{Soares} P.~S.,  {Watkinson} C.~A.,  {Cunnington} S.,   {Pourtsidou} A.,  2022,
  \mn@doi [\mnras] {10.1093/mnras/stab2594}, \href
  {https://ui.adsabs.harvard.edu/abs/2022MNRAS.510.5872S} {510, 5872}

\bibitem[\protect\citeauthoryear{{Spinelli}, {Carucci}, {Cunnington}, {Harper},
  {Irfan}, {Fonseca}, {Pourtsidou}  \& {Wolz}}{{Spinelli}
  et~al.}{2022}]{skaFGsim}
{Spinelli} M.,  {Carucci} I.~P.,  {Cunnington} S.,  {Harper} S.~E.,  {Irfan}
  M.~O.,  {Fonseca} J.,  {Pourtsidou} A.,   {Wolz} L.,  2022, \mn@doi [\mnras]
  {10.1093/mnras/stab3064}, \href
  {https://ui.adsabs.harvard.edu/abs/2022MNRAS.509.2048S} {509, 2048}

\bibitem[\protect\citeauthoryear{{Thorne}, {Dunkley}, {Alonso}  \&
  {N{\ae}ss}}{{Thorne} et~al.}{2017}]{pysm}
{Thorne} B.,  {Dunkley} J.,  {Alonso} D.,   {N{\ae}ss} S.,  2017, \mn@doi
  [\mnras] {10.1093/mnras/stx949}, \href
  {https://ui.adsabs.harvard.edu/abs/2017MNRAS.469.2821T} {469, 2821}

\bibitem[\protect\citeauthoryear{{Tingay} et~al.,}{{Tingay} et~al.}{2013}]{mwa}
{Tingay} S.~J.,  et~al., 2013, \mn@doi [\pasa] {10.1017/pasa.2012.007}, \href
  {https://ui.adsabs.harvard.edu/abs/2013PASA...30....7T} {30, e007}

\bibitem[\protect\citeauthoryear{{Waelkens}, {Jaffe}, {Reinecke}, {Kitaura}  \&
  {En{\ss}lin}}{{Waelkens} et~al.}{2009}]{magfield0}
{Waelkens} A.,  {Jaffe} T.,  {Reinecke} M.,  {Kitaura} F.~S.,   {En{\ss}lin}
  T.~A.,  2009, \mn@doi [\aap] {10.1051/0004-6361:200810564}, \href
  {https://ui.adsabs.harvard.edu/abs/2009A&A...495..697W} {495, 697}

\bibitem[\protect\citeauthoryear{{Wang} et~al.,}{{Wang} et~al.}{2021}]{wang}
{Wang} J.,  et~al., 2021, \mn@doi [\mnras] {10.1093/mnras/stab1365}, \href
  {https://ui.adsabs.harvard.edu/abs/2021MNRAS.505.3698W} {505, 3698}

\bibitem[\protect\citeauthoryear{{Wehus} et~al.,}{{Wehus} et~al.}{2017}]{wehus}
{Wehus} I.~K.,  et~al., 2017, \mn@doi [\aap] {10.1051/0004-6361/201525659},
  \href {https://ui.adsabs.harvard.edu/abs/2017A&A...597A.131W} {597, A131}

\bibitem[\protect\citeauthoryear{{Wolz}, {Abdalla}, {Blake}, {Shaw}, {Chapman}
  \& {Rawlings}}{{Wolz} et~al.}{2014}]{lauraFGsim}
{Wolz} L.,  {Abdalla} F.~B.,  {Blake} C.,  {Shaw} J.~R.,  {Chapman} E.,
  {Rawlings} S.,  2014. pp 3271--3283 (\mn@eprint {arXiv} {1310.8144}),
  \mn@doi{10.1093/mnras/stu792}

\bibitem[\protect\citeauthoryear{{Yohana}, {Ma}, {Li}, {Chen}  \&
  {Dai}}{{Yohana} et~al.}{2021}]{fastFGsim}
{Yohana} E.,  {Ma} Y.-Z.,  {Li} D.,  {Chen} X.,   {Dai} W.-M.,  2021, \mn@doi
  [\mnras] {10.1093/mnras/stab1197}, \href
  {https://ui.adsabs.harvard.edu/abs/2021MNRAS.504.5231Y} {504, 5231}

\bibitem[\protect\citeauthoryear{{Zhang}, {Bunn}, {Karakci}, {Korotkov},
  {Sutter}, {Timbie}, {Tucker}  \& {Wandelt}}{{Zhang} et~al.}{2016}]{intFGsim}
{Zhang} L.,  {Bunn} E.~F.,  {Karakci} A.,  {Korotkov} A.,  {Sutter} P.~M.,
  {Timbie} P.~T.,  {Tucker} G.~S.,   {Wandelt} B.~D.,  2016, \mn@doi [\apjs]
  {10.3847/0067-0049/222/1/3}, \href
  {https://ui.adsabs.harvard.edu/abs/2016ApJS..222....3Z} {222, 3}

\bibitem[\protect\citeauthoryear{{Zheng} et~al.,}{{Zheng} et~al.}{2017}]{gsm}
{Zheng} H.,  et~al., 2017, \mn@doi [\mnras] {10.1093/mnras/stw2525}, \href
  {https://ui.adsabs.harvard.edu/abs/2017MNRAS.464.3486Z} {464, 3486}

\bibitem[\protect\citeauthoryear{{Zonca}, {Singer}, {Lenz}, {Reinecke},
  {Rosset}, {Hivon}  \& {Gorski}}{{Zonca} et~al.}{2019}]{healpy}
{Zonca} A.,  {Singer} L.,  {Lenz} D.,  {Reinecke} M.,  {Rosset} C.,  {Hivon}
  E.,   {Gorski} K.,  2019, \mn@doi [The Journal of Open Source Software]
  {10.21105/joss.01298}, \href
  {https://ui.adsabs.harvard.edu/abs/2019JOSS....4.1298Z} {4, 1298}

\makeatother
\end{thebibliography}


\appendix
\section{Map zero-levels}
 \label{appendix:A}
 
A per-pixel spectral index map can be made from emission maps at two different frequencies. Observational maps, however, do not only contain spatially varying sky signal due to both diffuse and compact astrophysical sources; they contain contributions from constant background temperatures. A well-known example of one such contribution being the 2.725\,K from the cosmic microwave background. In \autoref{sec:tdata} we use the temperature-temperature linear regression method to identify the combination of these constant background temperatures for the CHIPASS data. The linear regression needs to be performed in a region with good dynamic range across both frequencies and, in particular, within a region dominated by one type of diffuse emission so as to ensure a straight line fit.
 
\begin{figure}
 \centering
  {\includegraphics[width=0.89\linewidth]{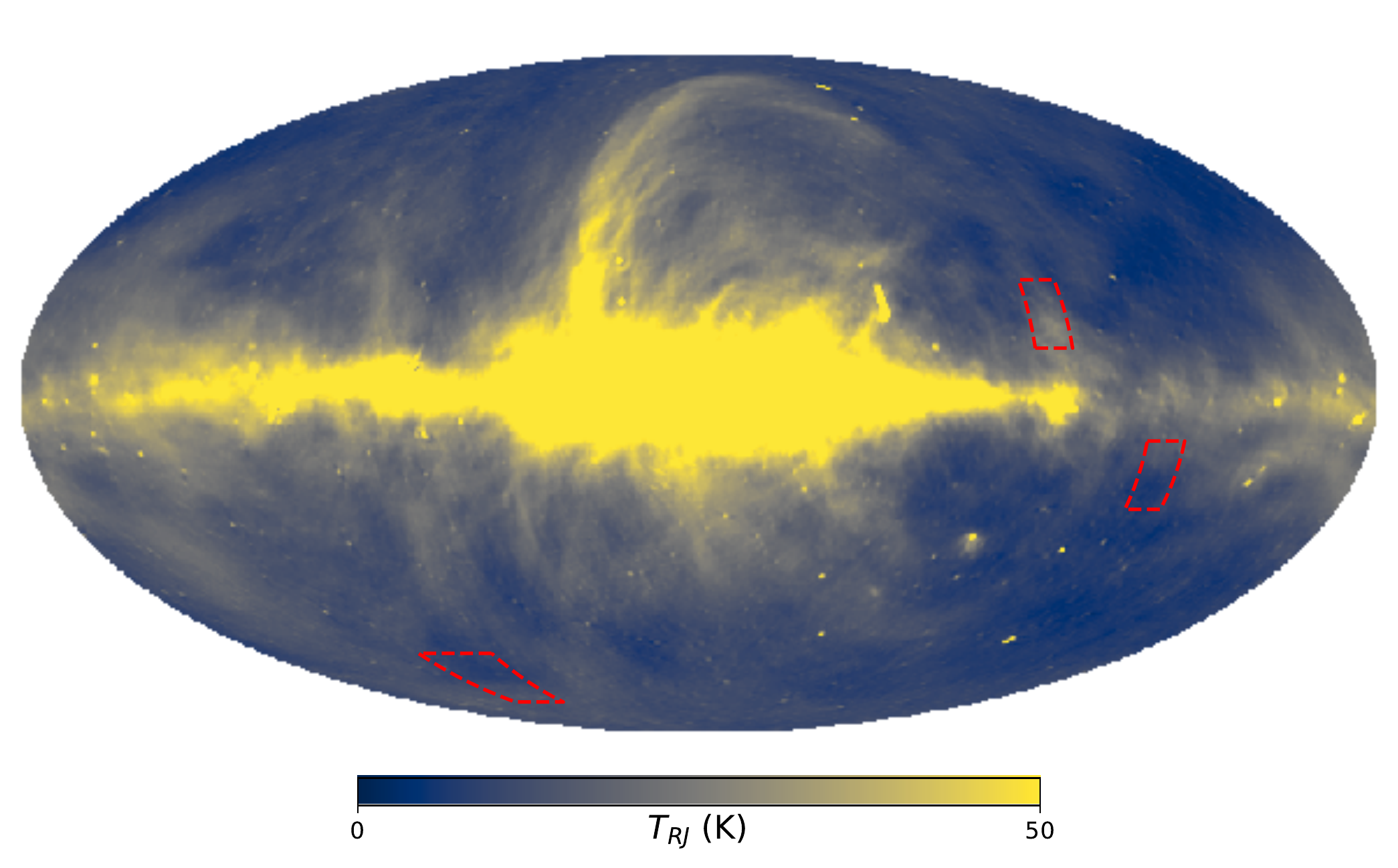}}\\
   {\includegraphics[width=0.89\linewidth]{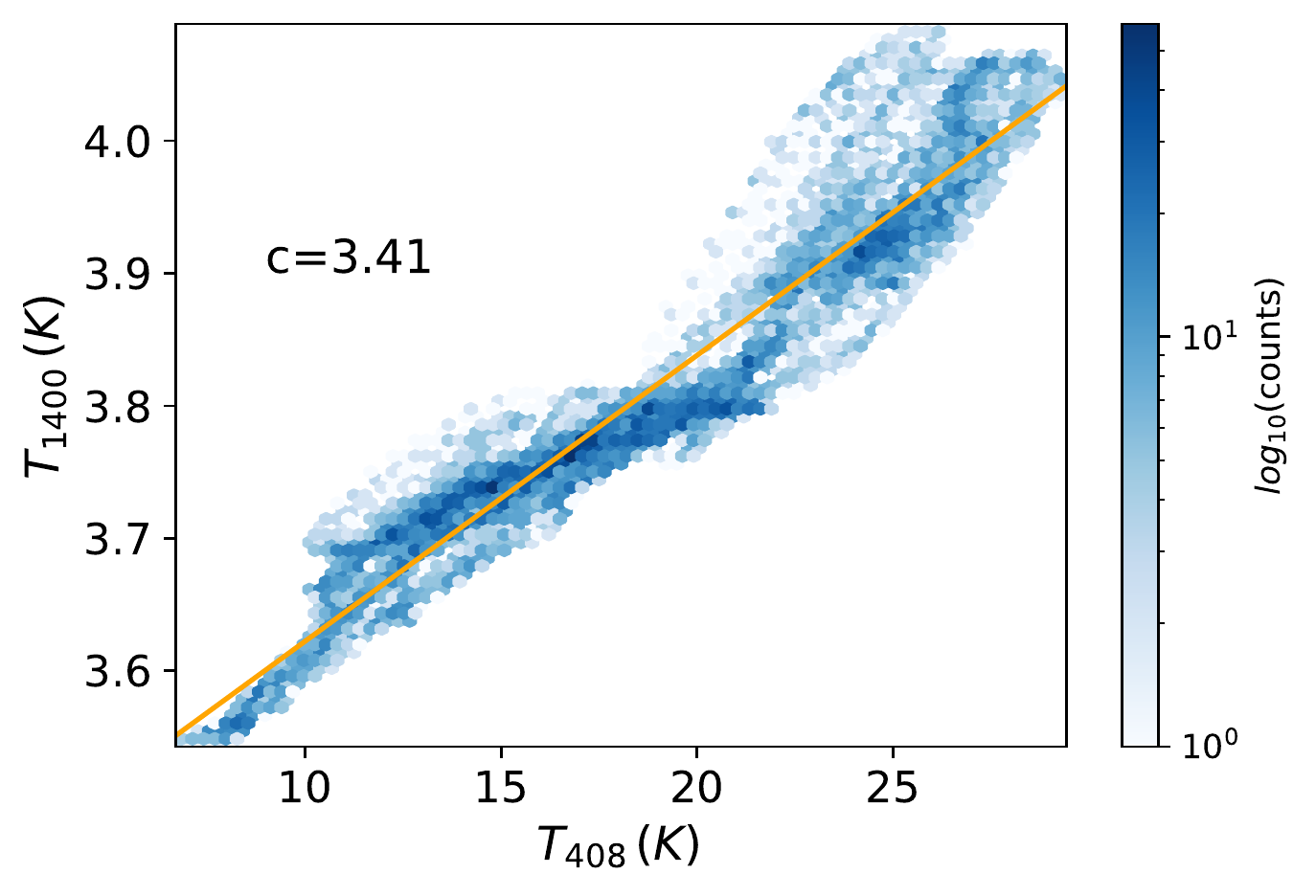}} \\
   {\includegraphics[width=0.89\linewidth]{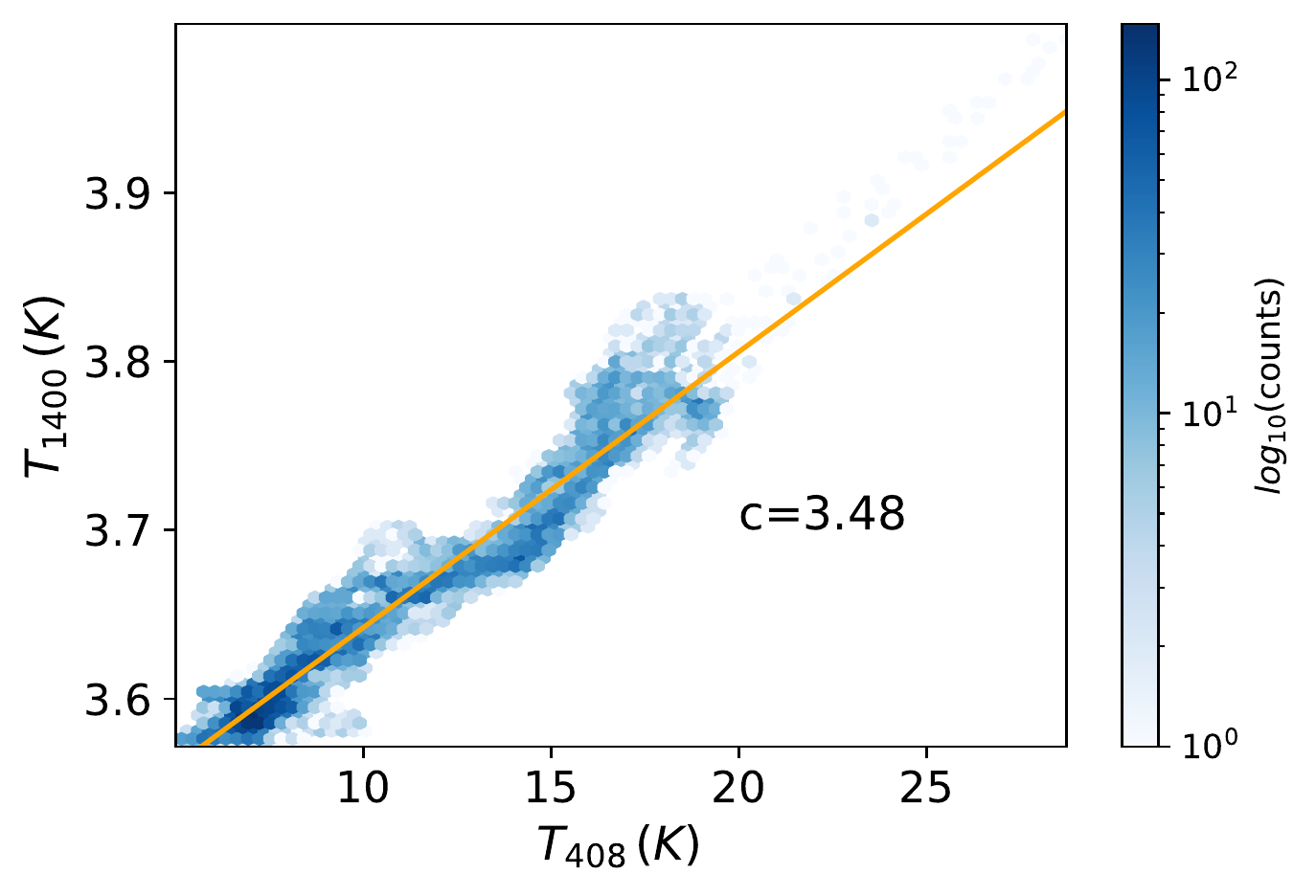}} \\
   {\includegraphics[width=0.89\linewidth]{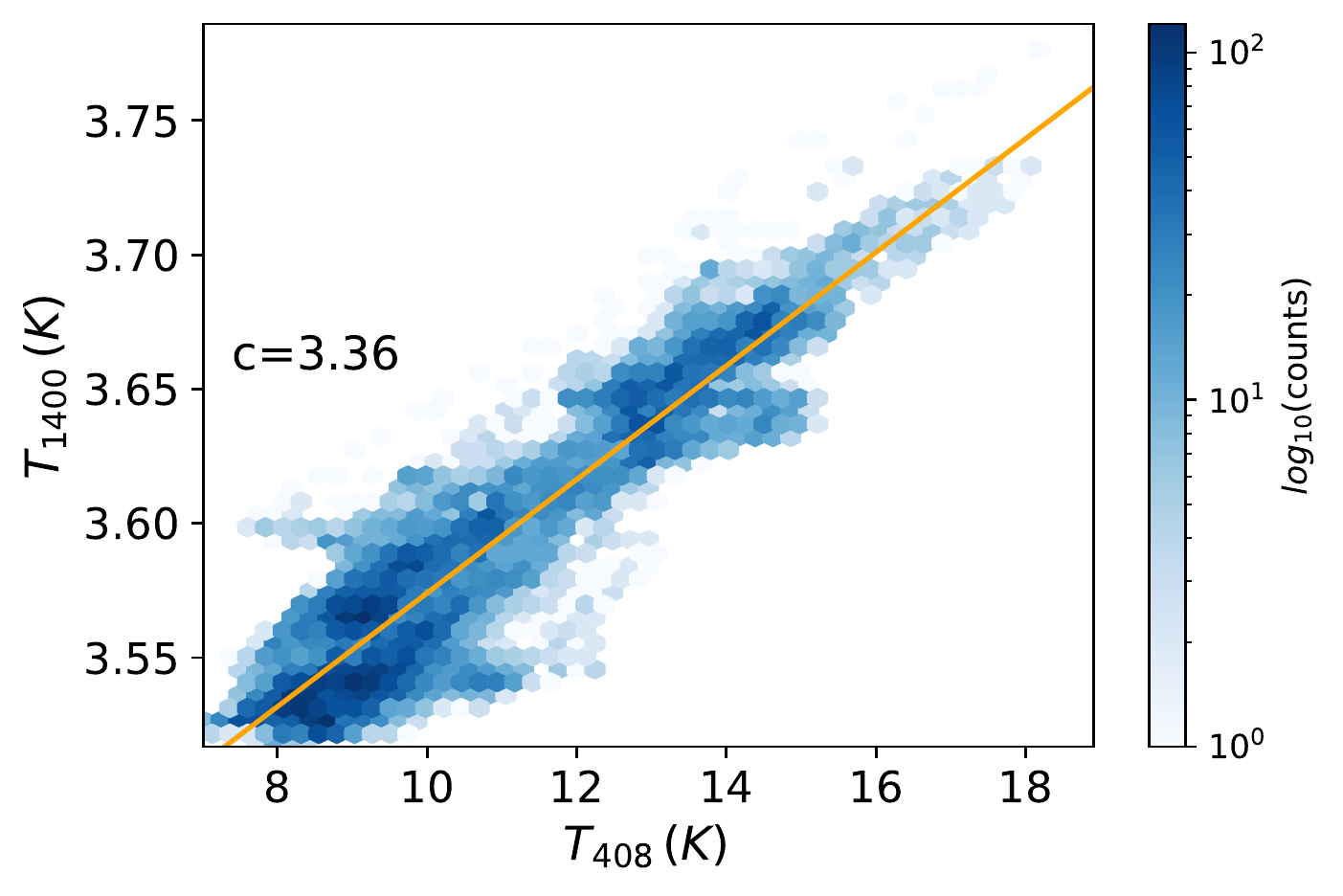}} \\
 \caption{{\it{Top:}} The three additional regions chosen to investigate the uncertainty on the 1.4\,GHz zero-level value. {\it{Second, third and fourth panel:}} Linear regression plots between 0.408 and 1.4\,GHz for each of the three regions (as read clockwise on the above map), the fitted zero-level is stated on each plot.}
 \label{fig:allinds}
   \end{figure}

We chose to use the North Polar Spur because this region contains bright, synchrotron-dominated emissions. In this appendix, we show the range of zero-levels obtained if one were to consider performing the linear regression in a different sky region. The top panel of \autoref{fig:allinds} highlights the three additional regions chosen with the following panels all showing the temperature-temperature linear regression plots between 0.408 and 1.4\,GHz for these regions. All the regions chosen have a high enough dynamic for a good linear fit (the $R^{2}$ value is greater than 0.9 for all the fits) but it can be seen that outside the North Polar Spur free-free emission is no longer negligible as they are spiral and horizontal patterns in the data which imply the presence of free-free emission at 1.4\,GHz which is not seen at 0.408\,GHz. As we believe the North Polar Spur to be the region most likely to be dominated by one emission mechanism we choose to use the zero-level of 3.21\,K but use the other three fits to inform our uncertainty estimate.
 
\begin{figure}
 \centering
  {\includegraphics[width=0.99\linewidth]{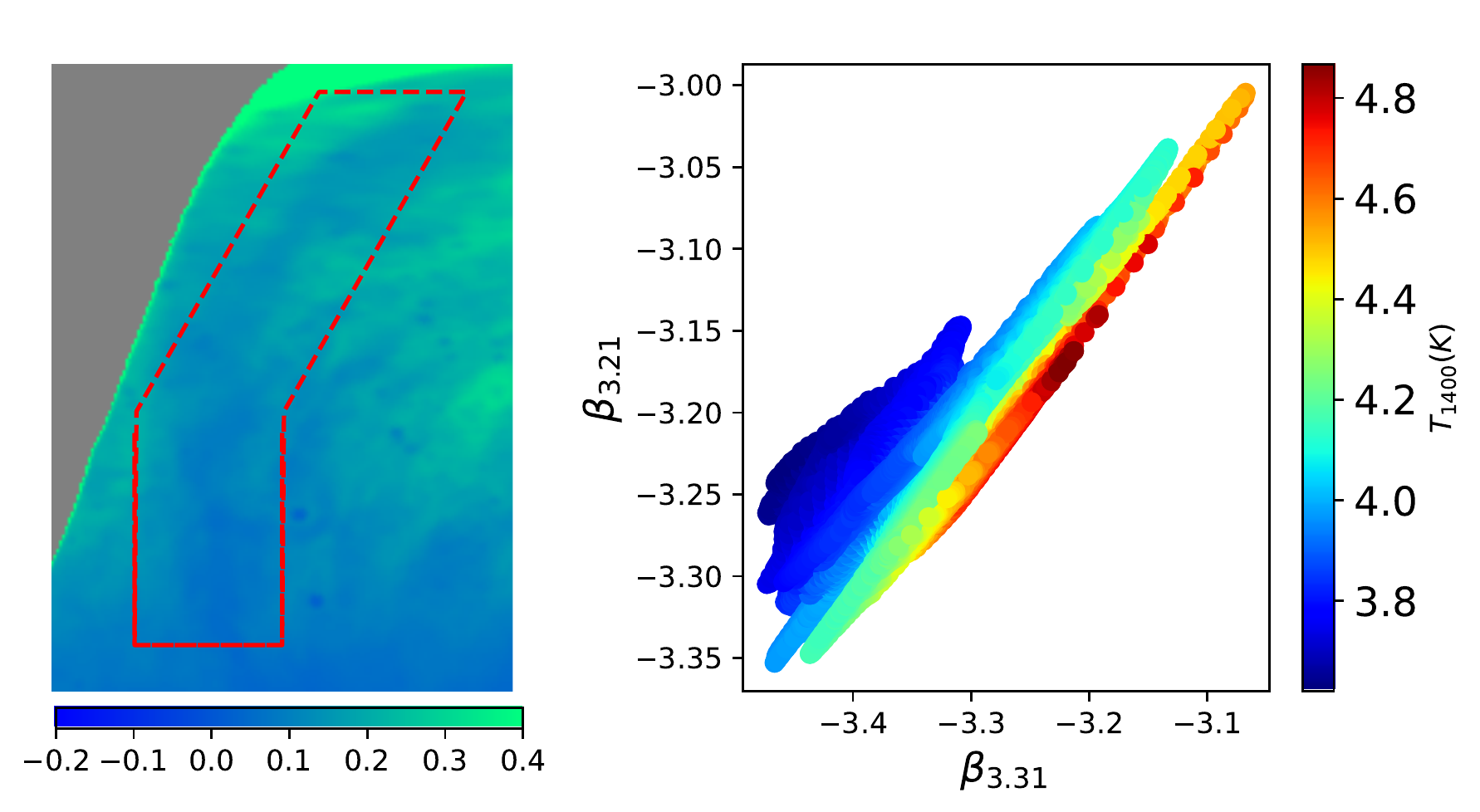}}
 \caption{{\it{Left: }}The difference in the 0.408-1.4\,GHz spectral index within the North Polar Spur region between one spectral index map made from CHIPASS data with a zero-level of 3.21\,K subtracted and the another spectral index map made from CHIPASS data with a zero-level of 3.31\,K subtracted. {\it{Right:}} The linear correlation between the two spectral index maps. The colour bar shows the 408\,MHz temperature at each pixel value.}
 \label{fig:indcom}
   \end{figure}

The standard deviation of the zero-level from these three fits (3.41\,K, 3.48\,K, 3.36\,K) as well as the North Polar Spur fit (3.21\,K) is 0.1\,K. We now consider two spectral index maps between 0.408 and 1.4\,GHz; one made from the CHIPASS data after a zero-level of 3.21\,K has been removed and one made using a zero-level of 3.31\,K for the CHIPASS data. The left panel of \autoref{fig:indcom} shows the difference between these two spectral index maps; not only can a difference in the mean spectral index level be seen, but there are also features within this map indicating that the spatial structure of the synchrotron spectral index is effected by the map zero-level values. In the right panel of \autoref{fig:indcom}, we plot the linear correlation between the two spectral index maps whilst simultaneously showing the temperature at 408\,MHz using the colour bar. For the pixels with the highest temperatures the correlation between the two spectral index maps is almost perfect, showing that for emission regions with temperatures significantly higher than the zero-level, small errors in the zero-level calculations make little difference to the spectral index spatial variations. However for the cooler portion of the North Polar Spur the change in zero-level results in a larger deviation in the spectral index spatial variation and weakens the strong correlation between the two maps. The correlation coefficient between the two spectral index maps made is 0.89; therefore we state an accuracy of 11 per cent for the spatial variations within our spectral index map.

\bsp	
\label{lastpage}
\end{document}